\documentclass[a4paper]{article}
    \usepackage{amsmath,amssymb}
     \usepackage{amsmath}
    \usepackage{mathrsfs}
    \usepackage{fancyhdr}
    \usepackage{eufrak}
    \usepackage{epigraph}
    \usepackage{latexsym}
    \usepackage{CJK}
    \usepackage{bm}
    \usepackage[english]{babel}
    \usepackage{blindtext}
    \usepackage[dvips]{graphicx}
    \begin{document}

\makeatletter
\renewcommand{\paragraph}{\@startsection{paragraph}{4}{0ex}%
   {-3.25ex plus -1ex minus -0.2ex}%
   {1.5ex plus 0.2ex}%
   {\normalfont\normalsize\bfseries}}
\makeatother

\stepcounter{secnumdepth}
\stepcounter{tocdepth}

    \newtheorem{defin}{Definition}[section]
    \newtheorem{theorem}[defin]{Theorem}
    \title{\textbf{On Measurement and Computation}}

    \author{Huimin Zheng\quad Haixing Hu\quad Nan Wu\quad Fangmin Song\\ \footnotesize{Department of Computer Science \& and Technology~}\\
\footnotesize{Nanjing University }\\\footnotesize{zhenghuimin555@126.com\quad fmsong@nju.edu.cn}
}
    \date{}
    \maketitle
    \begin{abstract}
    Inspired by the work of Feynman, Deutsch, We formally propose the theory of physical computation and accordingly, the physical complexity theory. To achieve this, a framework that could be used to evaluate almost all forms of computation making use of various physical mechanisms is established. Here, we focus on applying this to Quantum Computation. As a preliminary study on more general problems, some examples of other physical mechanisms are also discussed in this paper.  
    \end{abstract}

    \tableofcontents
    \section{Introduction}
    \subsection{Quantum Computation}
    The research of quantum computation has been lasting for about 30 years since R.Feynman proposed the concept of so-called `quantum computer' in 1982\cite{fey}. Founding out that there do exist some quantum systems which are suspected cannot be efficiently simulated by classical computers, early researchers naturally speculated that quantum mechanism itself may provide stunning power of computation. In order to strictly define what is `quantum computation', researchers introduced various new computational models, including Quantum Turing Machine \cite{ben} and Quantum Circuit Model\cite{acy}. However, at that time, no convincing evidence was discovered to support the conjecture that quantum mechanism can really be used to speed up the computation of some hard problems greatly.

    D.Deutsch found the first evidence that quantum computers may surpass the Turing Machine\cite{dj} in query. In fact, he constructed a special scene in which DTM has to query the oracle for $O(2^n)$ times to find the correct answer(for certain) in worst cases while QTM need just once query in all cases.

    One of the most remarkable results is quantum factorization, which is due to Peter Shor\cite{shor1,shor2}. The best classical algorithm for factorization so far has to run for~$O(\log^3 L)$~steps. However, Shor showed that one can use a family of quantum circuit, which contain~$O(\log^3 L)$~ gates and needs only~$O(L^2\log L\log\log L)$~operations to get the right answer.

    Grover's Algorithm~\cite{gro}~is another successful example of quantum algorithms. This algorithm can be used to search a database without structure. It is easy to proof that the time complexity of this problem for Turing Machine is $O(n)$. However, there does exist a quantum algorithm whose time complexity is $O(\sqrt{n})$. Since it has been proved that this is the optimal algorithm for all algorithms that considering quantum mechanics~\cite{nc}~, so the complexity of Grover's algorithm can be looked as the quantum complexity of this problem.

    One of the most important reason that why Quantum algorithms(especially Shor's algorithm) seem so interesting to many computer scientists is that their existence indicate a huge challenge to strong Church-Turing thesis, which
    states that \emph{Any model of computation can be simulated on a probabilistic Turing machine with at most a polynomial increase in the number of elementary operations required.}

    \subsection{Physical Computation}
    On the other hand, with the exciting research in quantum computation as well as other new paradigms of computations(e.g.DNA computation),
    the idea that we may just look physical processes as computations(not just the Turing Machine)was also developed. The seeds of this idea can be traced back to Feynman\cite{fey}, Deuthsch\cite{deu} and Pitowsky\cite{pit} et al.

    It is not very hard to understand and appreciate this idea, for at first glance, this point of view has at least three benefits:
    \begin{itemize}
      \item It can include the concept of classical algorithms easily, for an algorithm on Turing Machine(its physical implementation)can also be thought as a family of physical processes (and the corresponding measurement).
      \item We can try to solve some special kinds of problems with less time or space than the lower bound with respect to Turing machines.
      \item Being the ones which could be directly simulated, some physical methods can also enlighten us to design smart algorithms on Turing machines.
    \end{itemize}
    What's more, currently, it seems that we cannot exclude the possibility that there does exist a family of physical processes which can help us to calculate some problems which cannot be solved by a universal Turing Machine in principle.

    However, because of vagueness and extraordinary generality, the theory of so-called `physical computation' has a significant defeat yet.
    \begin{itemize}
      \item In many cases, people cannot decide how to define the resource for a `physical algorithm'. And as a result they cannot proof or even conjecture formally whether a `physical algorithm' is really superior to any algorithms of TMs with the same extensionality.
    \end{itemize}

    Note that the theory of quantum computation is almost free from such defeat, for researchers have completed the formal definition of the computational model of quantum computation in the early years. Roughly speaking, things tend to go wrong when:
    \begin{itemize}
      \item People adopt a design on which the physical postulates it depends is just an empirical one.
      \item More than one different systems of physical postulates are used.
    \end{itemize}
    \subsection{The structure of the article}
    The structure of this paper is as follows. In Sec. II we shall introduce two well-known models of quantum computation and the definition of complexity respectively. And after that we will formally establish the theoretical foundation of physical computation and propose the theory of physical computability in Sec. III. In the beginning of Sec. IV，we try to use the theory of physical computation to reanalyze the quantum algorithms. In the end of Sec. IV, we focus on the topic about how to construct problems which take advantage of quantum simulations.

    \section{Models of Quantum Computation}
    \subsection{Quantum Turing Machine}
    Quatnum Turing Machine was first introduced by Benioff\cite{ben} in 1980 and was developed by Deustch and Yao. The modern definitioin was given by Bernstein and Vazirani in 1997\cite{ber}.
    \begin{defin}\rm(Quantum Turing Machine, Bernstein~1997)
Let~$\tilde{C}$~be a set of complex nmber $\alpha$ satisfying: For each $\alpha$, there exists a polynomial time algorithm to compute the value of $Im(\alpha)$ and $Re(\alpha)$ close to $2^{-n}$ within the true value.

 A Quantum Turing Machine $M$ is defined as the triple $(\Sigma,Q,\delta)$, where $\Sigma$ is a fintie alphabet with an identified symbol $\#$ , $Q$~is a finite set of states with an identified initial state~$q_0$~and final state~$q_f\neq q_0$; $\delta$, the quantum transform function $\delta:Q\times\Sigma\rightarrow \tilde{C}^{\Sigma\times Q\times\{L,R\}}$. The QTM has a two-way infinite tape of cells indexed by $Z$, and a single read/write tape head that moves along the tape. We define configurations initial configurations and final configurations exactly as for DTMs.
 Let $S$ be the inner-product space of finite complex linear combinations of configurations of $M$ with the Euclidian norm. We call each element $phi\in S$ a superposition of $M$. The QTM $M$ defines a linear operator $U_M:S\rightarrow S$， called the time evolution operator of $M$ as follows: If $M$ starts in configurations $c$ with current state $p$ and scanned symbol $\sigma$. The after one step $M$ will be in superposition of configurations $\psi=\sum_i\alpha_ic_i$, where each non-zero $\alpha_i$ corresponds to a $\delta(p,\sigma,\tau,q,d)$, and $c_i$ is the new configuration that results from applying this transiton to $c$. Extending thi map to the entire space $S$ through linearity gives the linear time evolution operator $U_M$.
\end{defin}
\begin{defin}\rm
If $U_M$ can keep Euclidian norm, then we say $M$ is well deformed.
\end{defin}
\begin{theorem}\rm
If QTM is in the superposition $\psi=\sum_i\alpha_i c_i$ and is observed, the probability of the observer gets the configuration $c_i$ is $|\alpha_i|^2$, and then $M$ is in the state $\psi'=c_i$.
\end{theorem}
\begin{theorem}\rm A QTM is well-deformed if and only if its time evolution operator is unitary.
\end{theorem}
In QTM, the number of the read/write tape head moves during a computation is the cost of time.
\begin{theorem}\rm
There exists a universal QTM, which is polynomially equivalent to any QTMs.
\end{theorem}
    \subsection{Quantum Circuit Model}
    The first quantum circuit model was due to Deutsch. Then quantum circuit model was improved by Yao\cite{acy}, who also proved that for any QTM, there exists a uniform family of quantum circuit which is polynomially equivalent to that QTM.

    Not like QTM, quantum circuit model tends to describe an algorithm by using universal quantum gates and circuits without loops.
    Quantum circuit model does not need infinite many quantum gates, but finite many quantum gates which called the universal quantum gates.
    It has been proved that Hadamard Gate, phase gate, C-NOT Gate and $\pi/8$ Gate are universal.
    For any finite dimensional $U$ operators, we can always approach it effectively by means of a universal family of circuits $\mathscr{U}$, which only consists 4 gates above, i.e.
    $$\forall\varepsilon(\exists n\in\mathscr{U}),E(U,\tilde{U}_n)\equiv\max_{|\psi\rangle}\|(U-\tilde{U}_n)|\psi\rangle\|<\varepsilon$$

    The scale of a quantum circuit is defined as the number of the universal gates and the depth is defined as the longest path from input to output, if the gates is looked as a vertex.

    Both Quantum circuit model and QTM are important models of quantum computation. But we do not know whether they are the most natural models of quantum computation or do they fully take the advance of quantum mechanics, no matter in the theory of quantum computability and quantum computational complexity.

    \section{The Theory of Physical Computation}
    \subsection{Observer}
    Measurement is in terms of observer. Though there are many differences among people's opinions about the exact definition of human beings, we prudently assume that an observer is classical, that is, the observer will never get incompatible results during one measurement.

    In this article, we will never use the terminology such like `a observer of the observer', or in other words, by `observer' we always mean the last one outside the whole experiment.

    In order to unify various forms of results, we require that the observer only accept the symbols on a tape(just something like the one of TM) and also only use this to initialize an experiment.

    So we define the legal inputs and outputs as the elements in set $\Sigma^{+}$, where

    $$\Sigma=\{~0,1,*,.~\}$$ and $\Sigma^{+}$~\- the finite string composed by elements in~$\Sigma$.

    The concept of observer is crucial to our theory.

    \subsection{Physical States}
    We use (usually finite) attributes which may contribute to the computations to label the physical states. In addition, though may not be actually concerned in every computation, three fundamental quantities, namely, space, energy and mass are always included in a state for the sake of analysis of resource and complexity.

    So we have:
    $$\Omega\subset\{x_1\}^{A_1}\times\{x_2\}^{A_2}\times\dots\times\{x_n\}^{A_n}\times\{m\}^{\mathfrak{M}}\times\{s\}^{\mathfrak{S}}\times\{e\}^{\mathfrak{E}}$$

    Or more generally(Quantum)，
    $$\Omega\subset\{x_1\}^{A_1}\times\{x_2\}^{A_2}\times\dots\times\{x_n\}^{A_n}\times\{\mathbb{C}^m\}^{\mathfrak{M}}\times\{\mathbb{C}^s\}^{\mathfrak{S}}\times\{\mathbb{C}^e\}^{\mathfrak{E}}$$

    For simplicity, fundamental attributes are usually omitted, i.e.
    $$\Omega\subset\{x_1\}^{A_1}\times\{x_2\}^{A_2}\times\dots\times\{x_n\}^{A_n}$$

    For a certain attribute $A_i$, what really matters is its type which is constrained by its dimension. Note that dimensionless quantity(e.g.friction coefficient) can also be assigned to a null type. When a quantity is expressed by other quantities' combination, it's dimension type should be preserved, or rather, any equations should be dimensional balanced.

    For example:
    $$E^{[D:ML^2T^{-2}]}::=m^{[D:M]}g^{[D:LT^{-2}]}h^{[D:L]}=mgh^{[D:ML^2T^{-2}]}$$
    $$E^{[D:ML^2T^{-2}]}::=\frac{1}{2}(m)^{[D:M]}(v^2)^{[D:L^2T^{-2}]}=\frac{1}{2}(mv^2)^{[D:ML^2T^{-2}]}$$
    $$E^{[D:ML^2T^{-2}]}::=(m)^{[D:M]}(c^2)^{[D:L^2T^{-2}]}=(mc^2)^{[D:ML^2T^{-2}]}$$
    are all dimensional balanced.
    \subsection{Physical processes and the operation~$\circ$}
    Physical process on a state space $\Omega$ is a set of physical state whose elements are labeled by moment $t(t\in[0,T],T\in \mathbb{R}^{+})$.

    $$P\equiv\{~(T,\tilde{P})|~T\in \mathbb{R}^{+},\tilde{P}:[~0,T~]\to\Omega~\}$$

    If two physical processes on $\Omega$ satisfies $$(\pi_2P_1)(\pi_1P_1)=(\pi_2P_2)(0)$$ we can define operator $\circ:\{P\}\times \{P\}\to \{P\}$
    i.e.$$P_2\circ P_1=P_3$$satisfies:
     \begin{enumerate}
      \item $\pi_1P_3=\pi_1P_1+\pi_1P_2$
      \item if~$0\leq t\leq \pi_1 P_1$~，$(\pi_2P_3)(t)=(\pi_2P_1)(t)$
      \item if~$\pi_1P_1\leq t\leq \pi_1 P_1+\pi_1 P_2$~，$(\pi_2P_3)(t)=(\pi_2P_2)(t-\pi_1P_1)$
    \end{enumerate}
    For convenience, we introduce $\rhd P\lhd$ as the initial state of $P$, and $\lhd P\rhd$ the final state of $P$, i.e.
    $$\rhd P\lhd\equiv(\pi_2P)(0),\lhd P\rhd\equiv(\pi_2P)(\pi_1P)$$

    \subsection{Physical Operator and the operation of operator}
    Physical operator is a tuple whose first component is a state $x$ in $\Omega$ and the second component is a physical process whose initial state is $x$, i.e.
    $$O\equiv \{(x,P)|x\in\Omega,\rhd P\lhd=x\}$$
    In particular, a \emph{deterministic physical operator} $O$ means:$O$ is a physical operator, and
    $$\textrm{if~} \pi_2O(x_1)\neq\pi_2O(x_2),\textrm{then~} x_1\neq x_2$$
    if we only care about the effect the operator do to the initial state, we can look operator as a mapping in $\Omega$,i.e.$O:\Omega\to\Omega$.

    The operation between two deterministic physical operator is defined as follows(suppose $O_1,O_2$ are productive):
    $$O_3=O_2\circ O_1$$ which satisfies
    $$\forall x(O_3(x)\equiv O_2(\lhd O_1(x)\rhd)\circ O_1(x))$$

    In some more general cases, it is useful to talk about non-deterministic physical operators or random physical operators.
    A random physical operator $\tilde{O}$ contains the tuples which has the same initial states but different physical processes, i.e.
    $$\tilde{O}\equiv \{(x,P)|x\in\Omega,\rhd P\lhd=x\}.$$
    People cannot decide the output $\tilde{O}(x)$ just by the initial state $x\in\Omega$.

    Similarly, if just care about extensionality, we can look operator as a relationship on $\Omega$ i.e.$\tilde{O}:\Omega\times\Omega$

    We can also define operations between two non-deterministic operators, if some preconditions are satisfied. To do so, we first expand the definition of some symbols.
    $$O(x)\equiv\{P|\rhd P\lhd=x\}$$
    $$O(X)\equiv\{P|\rhd P\lhd\in X\subset \Omega\}$$
    $$\lhd O(x)\rhd\equiv\{y|y\in\Omega,\exists P\in O(x) s.t. \lhd P\rhd=y\}$$
    So $O_2\circ O_1$ (if they are productive) can be defined as：
    $$O_2\circ O_1(x)\equiv\{P_2\circ P_1|P_1\in O_1(x),P_2\in O_2(\lhd O_1(x)\rhd)\}$$

    Note that of all the processes created by random physical operators, their last states shall be exposed to outside world in the end by default.

    \subsection{Physical Computability}
    \subsubsection{Deterministic Physical Computation}
        \begin{defin}\rm (Deterministic Physical System)Deterministic Physical System $\mathscr{P}$ is a Five-Tuple
    $$\mathscr{P}\equiv(\Omega,\Sigma,\nabla,\mathcal{H},\Delta)$$where:
    \begin{itemize}
      \item $\Sigma=\{0,1,*,.\}$，$\Sigma^{+}$~is the collection of finite string formed by elements in~$\Sigma$~, $\Omega_{\Sigma^{+}}$~is the the set of physical implementation of $\Sigma^{+}$.
      \item $\Omega=\{\psi_i,i\in\Lambda\}$，~$\Omega\neq \Omega_{\Sigma^{+}}$，$\Lambda$~is an index set，~$\Omega$~is a set of distinguishable physical states(labeled by their attributes).
      \item $\nabla:\Omega_{\Sigma^{+}}\rightarrow \Omega$，\textbf{initialization operator}
      \item $\mathcal{H}:\Omega\rightarrow\Omega$~，\textbf{evolution operator}
      \item $\Delta:\Omega\rightarrow\Omega_{\Sigma^{+}}$，\textbf{Measurement operator}
    \end{itemize}
    \end{defin}
    Since Hilbert's 6th problem has not been solved yet, i.e. the whole theory of physics has not been axiomatized, we do not know that whether there exist some additional fundamental mathematical constraints should be included in this theory, though Beggs et al. have discussed the axioms of measurement based on Hempel's axioms\cite{beggs1}. Now, maybe the only restrictions here are the finiteness of the resource cost by a physical process and the finiteness of the attributes used to label a physical state set.

    As a result this system may looks looser than many classical computational models and may contains the ability to surpass all these models.
    We would like to let physicists to add more necessary restrictions into the system.

    Of course, when it comes to a specific branch of the physics, we can always know what is a legal states and processes. However, we wish to keep some freedom, i.e. to let the observer combine various axioms in physics so as to optimize the computations.

    \begin{defin}\rm (Partial Physical Computable Arithmetic Functions)For any partial arithmetic function~$f:\mathbb{N}\rightarrow \mathbb{N}$ is said to be \emph{partial physical computable}，if and only if there exists a Deterministic Physical System~$$\mathscr{P}\equiv(\Omega,\Sigma,\nabla,\mathcal{H},\Delta)$$ which satisfies：
    \begin{enumerate}
      \item[] If~$x\in dom(f)$, then,$$(\Delta\circ\mathcal{H}\circ\nabla)(x)\doteq f(x)$$
    \end{enumerate}
    \end{defin}

    Similarly, we can define \emph{Total Physically Computable Arithmetic Functions}

    \begin{defin}\rm  (Total Physically Computable Arithmetic Functions) For any total arithmetic function~$f:\mathbb{N}\rightarrow \mathbb{N}$ is said to be \emph{Total Physically Computable}, if and only if there exists a Deterministic Physical System~$(\Omega,\Sigma,\nabla,\mathcal{H},\Delta)$ which satisfies：
    \begin{enumerate}
      \item[] $\forall x\in\mathbb{N}$，we have:$$(\Delta\circ\mathcal{H}\circ\nabla)(x)\doteq f(x)$$
    \end{enumerate}
    \end{defin}

   In order to extend the definition of physical computability to non-arithmetic functions, we should take into consideration the precision of the measurement and computation. Therefore, we need a distance function to measure the precision of two values and define the computability as the ability of computing in any desired precision.

    \begin{defin}\rm (Partial Physically Computable Functions)Given a partial function~$f:A\rightarrow B$，and a metric~$\mathcal{D}:B\times B\rightarrow \mathbb{R}$，$f$~is said to be partial physically computable with respect to the metric ~$\mathcal{D}$~, if and only if for any~$\epsilon >0$~there exists~$(\Omega,\Sigma,\nabla,\mathcal{H},\Delta)$，s.t. for any~$x\in A$，we have：
    \begin{enumerate}
      \item[] if~$ x\in dom(f)$, $$\mathcal{D}\Big((\Delta\circ\mathcal{H}\circ\nabla)(x),f(x)\Big)<\epsilon$$
    \end{enumerate}
    \end{defin}

    Similarly, we can also define \emph{Total Physical Computable Functions}.

    \begin{defin}\rm (Total Physically Computable Functions)Given a total function~$f:A\rightarrow B$，and a metric~$\mathcal{D}:B\times B\rightarrow \mathbb{R}$，$f$~is said to be partial physically computable with respect to the metric ~$\mathcal{D}$~if and only if for any~$\epsilon >0$~there exists~$(\Omega,\Sigma,\nabla,\mathcal{H},\Delta)$, such that,
    \begin{enumerate}
      \item[] $\forall x\in A$ $$\mathcal{D}\Big((\Delta\circ\mathcal{H}\circ\nabla)(x),f(x)\Big)<\epsilon$$
    \end{enumerate}
    \end{defin}

     \subsubsection{Non-deterministic Physical Computation}
     On the other hand, many physical processes are considered to be non-deterministic, which enable us to implement so-called `randomized algorithms' and `quantum algorithms'. Our \emph{Probabilistic Physical System} is defined as follows.

    \begin{defin}\rm (Probabilistic Physical System)Probabilistic Physical System~$\mathscr{P}^{*}$~is a five-tuple:
    $$\mathscr{P}^{*}\equiv(\Omega,\Sigma,\nabla,\mathcal{H^{*}},\Delta)$$. where,
    \begin{itemize}
      \item $\Sigma=\{0,1,*,.\}$.
      \item $\Omega=\{\psi_i,i\in\Lambda\}$.
      \item $\nabla:\Omega_{\Sigma^{+}}\rightarrow \Omega$，which is also called \textbf{initialization operator}
      \item $\mathcal{H^{*}}:\Omega\times\Omega$~，which also called \textbf{evolution operator}, which is non-deterministic.
      \item $\Delta:\Omega\rightarrow\Omega_{\Sigma^{+}}$, which is also called \textbf{measurement operator}.
    \end{itemize}
    \end{defin}

    Non-deterministic does not necessarily cause probability, but let's convention that in this article we always discussed the randomness which has a probabilistic distribution.

    Definition of the computable functions by means of $\mathscr{P}^{*}$ is an analog to that of $\mathscr{P}$. As an example, we define Total Non-deterministic Physical Computable Functions.

    \begin{defin}\rm \textbf{(Total Non-deterministic Physical Computable Functions(Las Vegas))} For any total function~$f:\mathbb{N}\rightarrow \mathbb{N}$ is said to be total non-deterministic physical computable function, if and only if there exists a five-tuple~$$(\Omega,\Sigma,\nabla,\mathcal{H}^{*},\Delta)$$ s.t.
    \begin{enumerate}
      \item[]     $$\forall x((\Delta\circ \mathcal{H}^{*}\circ \nabla)(x)\doteq f(x))$$
    \end{enumerate}

    \end{defin}

    Because of randomness, for any identical inputs~$x$, the system may call different process to compute. The above definition is the counterpart of the definition of the so called Las Vegas algorithm.
    \begin{defin}\rm \textbf{(Total Non-deterministic Physical Computable Functions(Monte Carlo))}For any total function~$f:\mathbb{N}\rightarrow \mathbb{N}$ is said to be total non-deterministic physical computable, if and only if there exists~$$(\Omega,\Sigma,\nabla,\mathcal{H}^{*},\Delta)$$ s.t.
    \begin{enumerate}
      \item[] $\forall x\in\mathbb{N}$，$$\Pr\{Event(x)\text{occurs}\}>2/3$$
       where   $$Event(x)\equiv\Big((\Delta\circ\mathcal{H}^{*}\circ\nabla)(x)\doteq f(x)\Big)$$
    \end{enumerate}
    \end{defin}

    \subsection{Estimation of the  Complexity of Physical Resource}
    For the physical systems defined above, we can even ignore that whether there exists a physical mechanism in reality to implement it. Any functions which could be written as the composition of the three operators would be considered as computable(deterministic version).
   $$\begin{array}{ccc}
                   \Omega& \rightarrow \mathcal{H}&\Omega  \\
                  \nabla\uparrow &  & \Delta\downarrow \\
                 \Omega_{\Sigma^{+}}  & \rightarrow f & \Omega_{\Sigma^{+}}
                \end{array}
    $$

    But any experiments which implement a certain system will cost resource. We will focus on four kinds of resource, namely, time, space, energy and mass.

    One can also include 7 fundamental sorts of attributes considering 7 dimensions in SI(Syst\'{e}me international d'unit\'{e}s). However, actually our decision is not a totally empirical one. In fact, many differential equations which are established to describe various phenomena involving these attributes are always related to energy, time, and space. After properly choosing unit to make all the constant into 1, we can just rewrite these dimensions in the forms of the combinations of 4 fundamental dimensions, which enable us to continue to use the 4 attributes to represent the increase of the resource. The counterexample may occur only when $m$ or more than $m$ new attributes appear in $n$ equations and $m>n$, however these cases tend to be unlikely to happen, if these equations(theory) we discussed are assumed to be `enough' for some phenomena in nature.

    \begin{defin}\rm(\textbf{~Metric for Resource~}) The resource of a physical process~$\mathfrak{R}$ includes:
    \begin{itemize}
      \item $\mathfrak{T}$:~The (expectation of the)total time the whole process consumed;
      \item $\mathfrak{S}$:~The maximum of (the expectation of)the space the whole process consumed;
      \item $\mathfrak{M}$:~The maximum of (the expectation of)the mass the whole process consumed;
      \item $\mathfrak{E}$:~The maximum of (the expectation of)the energy the whole process consumed.
    \end{itemize}
    and~$\mathfrak{R}\equiv(\mathfrak{T},\mathfrak{S},\mathfrak{M},\mathfrak{E})$
    \end{defin}
    In the above definitions, the metric of them could be selected as the common ones.
    Today, most physicists tends to believe that mass and energy are not independent, neither do time and space. But for convenience, we still focus the primitive forms of resource, for actually we don't care about the independence here.

    In the above definitions, we don't talk about the potential possibility that even time could be reused.

    We convention that the resource is with respect to an inertial system, i.e. the observers obtain their results when they are in an inertial system to the system running the `algorithms', so as to rule out the paradoxes because of the theory of relativity.

    In many cases, we just cannot get a infinite precise estimation about the resource, but for our purpose, we actually do not need such things.
    Of course, there may exist some cases when we could not get an estimation without any promise of any precision, however, we will not use such processes to construct our implementation.

    Suppose the projections of the fundamental attributes(resource) are~$\pi_{\mathfrak{M}}(S)$, $\pi_{\mathfrak{S}}(S)$ and $\pi_{\mathfrak{E}}(S)$
    Then the resource a physical process consumed is:
    $$\begin{array}{lll}
        \mathfrak{T}P & \equiv & \pi_1 P \\
        \mathfrak{M}P & \equiv & \max\{\pi_{\mathfrak{M}}(S),S\in \mathrm{Ran}(\pi_2P)\} \\
        \mathfrak{S}P & \equiv & \max\{\pi_{\mathfrak{S}}(S),S\in \mathrm{Ran}(\pi_2P)\} \\
        \mathfrak{E}P & \equiv & \max\{\pi_{\mathfrak{E}}(S),S\in \mathrm{Ran}(\pi_2P)\}
      \end{array}
    $$
    In general cases, when we have to discuss the process of superposition, the resource can be defined as:
    $$\begin{array}{lll}
        \mathfrak{T}P & \equiv & \pi_1 P \\
        \mathfrak{M}P & \equiv & \max\{\mathrm{E}[\pi_{\mathfrak{M}}(S)],S\in \mathrm{Ran}(\pi_2P)\} \\
        \mathfrak{S}P & \equiv & \max\{\mathrm{E}[\pi_{\mathfrak{S}}(S)],S\in \mathrm{Ran}(\pi_2P)\} \\
        \mathfrak{E}P & \equiv & \max\{\mathrm{E}[\pi_{\mathfrak{E}}(S)],S\in \mathrm{Ran}(\pi_2P)\}
      \end{array}
    $$
    So it is easy to see that
    $$\begin{array}{lll}
        \mathfrak{T}P_2\circ P_1 & = & \pi_1P_1+\pi_1P_2 \\
        \mathfrak{M}P_2\circ P_1 & = & \max\{\mathfrak{M}P_1,\mathfrak{M}P_2\} \\
        \mathfrak{S}P_2\circ P_1 & = & \max\{\mathfrak{M}P_1,\mathfrak{M}P_2\} \\
        \mathfrak{E}P_2\circ P_1 & = & \max\{\mathfrak{M}P_1,\mathfrak{M}P_2\}
      \end{array}
    $$
    According to the definition above, the resource of a non-deterministic physical operator $O$ which is initialized by $x\in\Omega$ should be defined as:
    $$\begin{array}{lll}
        \mathfrak{T}O(x) & \equiv & \mathrm{E}[\mathfrak{T}O_i(x)] ,O_i(x)\in O(x)\\
        \mathfrak{M}O(x) & \equiv & \mathrm{E}[\mathfrak{M}O_i(x)] ,O_i(x)\in O(x) \\
        \mathfrak{S}O(x) & \equiv & \mathrm{E}[\mathfrak{S}O_i(x)] ,O_i(x)\in O(x) \\
        \mathfrak{E}O(x) & \equiv & \mathrm{E}[\mathfrak{E}O_i(x)] ,O_i(x)\in O(x) \\
      \end{array}
    $$
    So for operators' operation, we have:
    $$\begin{array}{lll}
        \mathfrak{T}O_2\circ O_1(x) & = & \mathfrak{T}O_1(x)+\mathfrak{T}O_2(\lhd O_1(x)\rhd)\\
        \mathfrak{M}O_2\circ O_1(x) & = & \max\{\mathfrak{M}O_1(x),\mathfrak{M}O_2(\lhd O_1(x)\rhd)\} \\
        \mathfrak{S}O_2\circ O_1(x) & = & \max\{\mathfrak{S}O_1(x),\mathfrak{S}O_2(\lhd O_1(x)\rhd)\} \\
        \mathfrak{E}O_2\circ O_1(x) & = & \max\{\mathfrak{E}O_1(x),\mathfrak{E}O_2(\lhd O_1(x)\rhd)\} \\
      \end{array}
    $$

    \subsubsection{Framework for the Complexity with respect to General Physical Resource}
    \begin{defin}\rm\textbf{(Resource(deterministic))}A resource  the physical process which complete the whole computation consumed $\mathfrak{R}_{\mathscr{P}}$ including:
    $$\begin{array}{lll}
        \mathfrak{T}_{\mathscr{P}} & \equiv & \mathfrak{T}((\Delta\circ \mathcal{H}\circ \nabla)(x))\\
                     &   & \\
        \mathfrak{S}_{\mathscr{P}} & \equiv & \mathfrak{S}((\Delta\circ \mathcal{H}\circ \nabla)(x))\\
                    &   & \\
        \mathfrak{M}_{\mathscr{P}} & \equiv & \mathfrak{M}((\Delta\circ \mathcal{H}\circ \nabla)(x))\\
                    &   & \\
        \mathfrak{E}_{\mathscr{P}} & \equiv & \mathfrak{E}((\Delta\circ \mathcal{H}\circ \nabla)(x))\\
      \end{array}
    $$
i.e.~$\mathfrak{R}_{\mathscr{P}}\equiv(\mathfrak{T_{\mathscr{P}}},\mathfrak{S}_{\mathscr{P}},\mathfrak{M}_{\mathscr{P}},\mathfrak{E}_{\mathscr{P}})$
    \end{defin}
    \begin{defin}\rm\textbf{(Resource(Las Vegas))}A resource  the physical process which complete the whole computation consumed $\mathfrak{R}_{\mathscr{P}}$ including:
    $$\begin{array}{lll}
        \mathfrak{T}_{\mathscr{P}} & \equiv & \mathfrak{T}((\Delta\circ \mathcal{H}\circ \nabla)(x))\\
                     &   & \\
        \mathfrak{S}_{\mathscr{P}} & \equiv & \mathfrak{S}((\Delta\circ \mathcal{H}\circ \nabla)(x))\\
                    &   & \\
        \mathfrak{M}_{\mathscr{P}} & \equiv & \mathfrak{M}((\Delta\circ \mathcal{H}\circ \nabla)(x))\\
                    &   & \\
        \mathfrak{E}_{\mathscr{P}} & \equiv & \mathfrak{E}((\Delta\circ \mathcal{H}\circ \nabla)(x))\\
      \end{array}
    $$
i.e.~$\mathfrak{R}_{\mathscr{P}}\equiv(\mathfrak{T_{\mathscr{P}}},\mathfrak{S}_{\mathscr{P}},\mathfrak{M}_{\mathscr{P}},\mathfrak{E}_{\mathscr{P}})$
    \end{defin}

        \begin{defin}\rm\textbf{(Resource(Monte Carlo))}A resource  the physical process which complete the whole computation consumed $\mathfrak{R}_{\mathscr{P}}$ including:
    $$\begin{array}{lll}
        \mathfrak{T}_{\mathscr{P}} & \equiv & \mathfrak{T}((\Delta\circ \mathcal{H}\circ \nabla)(x))\\
                     &   & \\
        \mathfrak{S}_{\mathscr{P}} & \equiv & \mathfrak{S}((\Delta\circ \mathcal{H}\circ \nabla)(x))\\
                    &   & \\
        \mathfrak{M}_{\mathscr{P}} & \equiv & \mathfrak{M}((\Delta\circ \mathcal{H}\circ \nabla)(x))\\
                    &   & \\
        \mathfrak{E}_{\mathscr{P}} & \equiv & \mathfrak{E}((\Delta\circ \mathcal{H}\circ \nabla)(x))\\
      \end{array}
    $$
i.e.~$\mathfrak{R}_{\mathscr{P}}\equiv(\mathfrak{T_{\mathscr{P}}},\mathfrak{S}_{\mathscr{P}},\mathfrak{M}_{\mathscr{P}},\mathfrak{E}_{\mathscr{P}})$
    \end{defin}
     The corresponding concept of complexity should be defined as the resource consumed with respect to the length of the input.

    \begin{defin}\rm\textbf{Complexity}
    The complexity of a kind of resource is a function of the length of the input $x$ $n=\lceil\log x\rceil$

    $$\begin{array}{lll}
        \mathrm{Conplexity}_t(n) & = & \max\{\mathfrak{T}_{\mathscr{P}(x)}|n-1\leq\log x\leq n\} \\
        \mathrm{Conplexity}_m(n) & = & \max\{\mathfrak{M}_{\mathscr{P}(x)}|n-1\leq\log x\leq n\} \\
        \mathrm{Conplexity}_s(n) & = & \max\{\mathfrak{S}_{\mathscr{P}(x)}|n-1\leq\log x\leq n\} \\
        \mathrm{Conplexity}_e(n) & = & \max\{\mathfrak{E}_{\mathscr{P}(x)}|n-1\leq\log x\leq n\}
      \end{array}
    $$

    \end{defin}
    Note: \emph{actually, the finiteness of resource cost is a necessary precondition for all ``uniform'' physical computation models. Another necessary precondition is that all the physical attributes should be at some trivial states(very easily to be constructed, e.g. $0^{\circ}\mathrm{C}$, $0$m/s) before the experiments. If these requirement are not satisfied, the model will be a non-uniform one. This case is also discussed in details by Beggs et al.\cite{beggs1},\cite{beggs4}}

    \subsubsection{Some Common Examples}
    It is interesting to find some new methods to compute problems without the help of universal Turing Machines. Through the ages, people have found a lot of such examples, the most famous of them are:
     \begin{itemize}
      \item Measure the volume of an object by putting it into the water;
      \item Obtain the centroid of an object by two suspension method;
      \item Compute function sine by analog circuit;
      \item Decide the path of minimum cost using Fermat's Principle;
      \item Calculate the mean of numbers by the second Law of Thermodynamics.
    \end{itemize}

    Actually, we can give even more similar examples:
    \begin{itemize}
    \item
    By making use of resonance, we can easily find the desired tuning fork from a heap of them. Otherwise, we have to look up the label of them one by one and even have to compute the frequency one by one if there is no labels on them.
      \item
      We can compute the square root of an given number $x$ by the law of free fall. Prepare a vacuum tube T of length $x$ and let it stand vertically, then let an object o which is small enough fall. Get the time $t$ when it touch the bottom, and we have $\sqrt{x}=t/c$, where $c=(2/g)^{1/2}$.
      \item We can sort a series of numbers through dangling poises by strings, where the strings satisfies Hooke's law. Given an array of numbers $\{x_i\}$ construct or find poises whose mass is just $x_i$, then dangle them by strings with the same stiffness coefficient. When the system is stable, the position of the poises with respect to their weight just indicate the relationship desired.
    \end{itemize}

    However, it is hard for us to estimate the cost of the methods above just after we describe them informally. So we select a part of them to analyze next.

    \textbf{Conventions:} $x$ is the representation of number in digits, $[x]$ is the value of $x$, $[x]^A$ means the attribute $A$ has the value $x$. $[x]^{\Sigma^*}$ means the representation of quantity $x$ though not on the tape.

    \paragraph{Mean of Three Numbers}
    Given three numbers, compute the mean of them making use of law of thermodynamics. This idea comes from Pitowsky~\cite{pit}.

    The strict description of the problem: Given three numbers $x_1$, $x_2$, $x_3\in [~0,100~]$, compute $$\bar{x}=\frac{(x_1+x_2+x_3)}{3}\text{\quad(Accurate to two decimal places)}.$$
    Pitowsky suggests that since all of the three numbers less than 100 and bigger than zero, note that the freezing point of water is 0~$C^{\circ}$ and the boiling point of water is 100~$C^{\circ}$ under the one standard air pressure, So for each number $x_i$, we can prepare the corresponding water of volume $V$ and temperature of $x_i C^{\circ}$. And then pour the water of three vessels into a bigger one, whose volume is $V'(V'>3V)$, and wait. After the water arrived at the balance point, measure the temperature. Of course, we assume that during the whole procedure, no calory is lose.

    Apparently, the physical state the method above deal with is the temperature of water, so we have $$\Omega=\{\vec{t}~|t_i\in [0,100],i=1,2,3\}^T.$$ on the other hand, we suppose the water is heat up from 0~$C^{\circ}$, i.e. the initial state of the experiment is $([0]^T,[0]^T,[0]^T)$.

    Therefore, the process could be depicted as following:

    $\nabla:\Sigma^{+}\rightarrow \Omega$, heat up the water to the desired temperature
    $$\nabla (x_1,x_2,x_3)=([x_1]^T,[x_2]^T,[x_3]^T)$$

    $\mathcal{H}: \Omega\rightarrow \Omega$, admixture the water of different temperature, the second law of thermodynamics is used
    $$\mathcal{H}([x_1]^T,[x_2]^T,[x_3]^T)=([\bar{x}]^T,[\bar{x}]^T,[\bar{x}]^T)$$

    $\Delta: \Omega\rightarrow \Sigma^{+}$, measure the temperature of the water
    $$\Delta([\bar{x}]^T,[\bar{x}]^T,[\bar{x}]^T)=\bar{x}$$

    For this problem, since the precision is finite, and there are only constant (three) numbers and the numbers are bounded, we can easily deduct that $\mathfrak{R}_{\mathscr{P}}$ is a constant. As a matter of fact, for Turing Machine, we can also find a constant resource costing algorithm which is just looking up a finite list to solve the problem.

    \paragraph{Sorting Without Repeat}

    Description of the Problem:
              \begin{description}
                \item[Input:] Finite number series of length $n$: $$A=\{x_i|x_i\in \mathbb{Z^+}\cap [0,M](0\leq i\leq n)\};$$
                \item[Output:] Finite number series of length $m$: $$B=\{x_j|x_j\in A(0\leq j\leq m)\},$$ s.t. if~$j_1<j_2$~then~$x_{j_1}<x_{j_2}$~.
              \end{description}
    Our plan is: for the given series, select a series of poises of length $n$, s.t. the mass of the $i$th poise is equivalent to the $i$th number. Dangling the poises from right to left by strings, whose restoring coefficient are $k$. Wait until the system is stationary, open the parallel light source and measure the projection onto the vertical ruler at the right end. The measurement could be done by machines and present the results onto the tape for observer. For Example, we can embed some photoconductive diodes in the ruler by graduations, diodes who is not triggered should be read.

    The physical state the method is primarily concerned with is the mass of poise $M$, the horizontal positions of the poises $X$ and the vertical ones $Y$, the projections $Y'$ and the boole value~$B$ indicating which diodes is triggered, i.e.

    $$\Omega\equiv\oplus^n_i\{m_i\}^M\times\{x_i\}^X\times\{y_i\}^Y\times\oplus_j\{(j,B_j)\}^{Y'\times B}$$

    So we have

     $\nabla:\Sigma^{+}\rightarrow \Omega$ (Select poises)
    $$\nabla (\oplus^n_i x_i)=(\oplus_i[x_i]^M[i]^X[0]^Y\oplus^M_j[(j,0)]^{Y'\times B})$$

    $\mathcal{H}: \Omega\rightarrow \Omega$ (Dangle poises)
    $$\mathcal{H}\left(\begin{array}{l}
                   \oplus^n_i[x_i]^M \\
                   {[i]^X} \\
                   {[0]^Y} \\
                   \oplus^M_j[(j,0)]^{Y'\times B}
                 \end{array}\right)=\left(
                                      \begin{array}{l}
                                        \oplus^n_i[x_i]^M \\
                                        {[i]^X} \\
                                        {[[x_i]g/k]^Y} \\
                                        \oplus^M_j[(j,0)]^{Y'\times B} \\
                                      \end{array}
                                    \right)
    $$

    $\mathcal{H'}: \Omega\rightarrow \Omega$(Open the parallel light)
    $$\mathcal{H'}\left(
                   \begin{array}{l}
                    \oplus^n_i[x_i]^M \\
                     {[i]^X} \\
                     {[[x_i]g/k]^Y} \\
                     \oplus_j[(j,0)]^{Y'\times B} \\
                   \end{array}
                 \right)
    =\left(
       \begin{array}{l}
         \oplus^n_i[x_j]^M \\
         {[j]^X} \\
         {[[x_j]g/k]^Y} \\
         \oplus^M_j[(j,\epsilon_A(j))]^{Y'\times B} \\
       \end{array}
     \right)
   $$

    $\Delta: \Omega\rightarrow \Sigma^{+}$(read the projection)
    $$\Delta\left(
              \begin{array}{c}
                \oplus^n_i([x_j]^M \\
                {[j]^X} \\
                {[[x_j]g/k]^Y} \\
                \oplus^M_j[(j,\epsilon_A(j))]^{Y'\times B}) \\
              \end{array}
            \right)=\oplus_{j'}(x_{j'})
    $$
    satisfies if~$j_1<j_2$~then$$[x_{j_1}]<[x_{j_2}]$$

    Considering the ideal implementation, we conclude that the $\mathfrak{R}_{\mathscr{P}}$ is linear, which is superior to Turing Machines using comparisons, for the complexity for them is proofed to be $O(n\log n)$. However, there does exist Turing Machine, which is not based on comparisons, also has a linear time cost.

    Note that if the number series is boundless, the complexity of the method above will be exponential. This is the common defeat of most analog computers.

    \paragraph{Volume of irregular shape}
    For this issue, we shall restrict the range of the saying `irregular' so as to rule out the objects with infinite length of description. So actually, we tend to discuss a subset of the set of all cases.

    Description of the problem:
    \begin{description}
      \item[Inputs:] point series of length $n$:$(x_i,y_i)(1<i<n)$, satisfies $c+r\leq x_i\leq a-c-r,c+r\leq y_i\leq b-c-r$
      \item[Outputs:] The volume of the box of length $a$ and width $b$ and height $h_0$, not including the series of cylinders(radius:~$r$~height:~$h_0$) which are induced by the series of points.
    \end{description}
    Our plan is simple. Assume we have a box of material of dense $\rho$, and a punch to extract circles from it. Then we measure the mass of the rest then divide it by its dense or just put it into water.
     $\nabla:\Sigma^{+}\rightarrow \Omega$
    $$\nabla(\oplus^n_{i=1}a(x_i,y_i))=[\rho h_0(A-m(\cup^n_{i=1}c_i))]^M[0]^{\Sigma'}$$

    $\mathcal{H}_1:\Omega\rightarrow\Omega$
    $$\mathcal{H}_1[\rho h_0(A-m(\cup^n_{i=1}c_i))]^M[0]^{\Sigma'}$$$$=[\rho h_0(A-m(\cup^n_{i=1}c_i))]^M[\rho h_0(A-m(\cup^n_{i=1}c_i))]^{\Sigma'}$$

    $\mathcal{H}_2:\Omega\rightarrow\Omega$
    $$\mathcal{H}_2
    [\rho h_0(A-m(\cup^n_{i=1}c_i))]^M[\rho h_0(A-m(\cup^n_{i=1}c_i))]^{\Sigma'}$$$$=[\rho h_0(A-m(\cup^n_{i=1}c_i))]^M[ h_0(A-m(\cup^n_{i=1}c_i))]^{\Sigma'}$$

    $\Delta:\Omega\rightarrow\Sigma^{+}$
    $$\Delta[\rho h_0(A-m(\cup^n_{i=1}c_i))]^M[ h_0(A-m(\cup^n_{i=1}c_i))]^{\Sigma'}=h_0(A-m(\cup^n_{i=1}c_i))$$

    Apparently the resource complexity for this method is linear with respect to the number of the points. However, because most people think that we cannot do infinitely measurement during one experiment, this method can only provide the result of finite precision. This is a good news to Turing Machines because this implies there exists a Turing Machine which is almost equivalently efficient.

    This may be astonish to someone, who may thought that a TM should at least solve the equations first. However, because of the finite precision, Turing Machine can just split the object into lattice and use the so-called scan-line algorithm to find the answer.

    \paragraph{The centroid of Irregular Shape}
    Just as the last example, we restrict our topic into the same subsets of all cases.

    Description of Problem:
    \begin{description}
      \item[Inputs:] point series of length $n$:$(x_i,y_i)(1<i<n)$, satisfies $c+r\leq x_i\leq a-c-r,c+r\leq y_i\leq b-c-r$
      \item[Outputs:] The centroid of the box of length $a$ and width $b$ and height $h_0$, not including the series of cylinders which is induced by the series of points.
    \end{description}

    The method we suggest is similar to the last one, the difference of them is that this time we will record some points.

    $\nabla:\Sigma^{+}\rightarrow \Omega$
    $$\nabla(\oplus^n_{i=1}a(x_i,y_i))=[\rho h_0(A-m(\cup^n_{i=1}c_i))]^M[0]^{\Sigma}[0]^{\Sigma'}[0]^{\Sigma''}$$

    $\mathcal{H}_1:\Omega\rightarrow\Omega$~(Suspend the box by $V_0$)
    $$\mathcal{H}_1[\rho h_0(A-m(\cup^n_{i=1}c_i))]^M[0]^{\Sigma}[0]^{\Sigma'}[0]^{\Sigma''}$$$$=[\rho h_0(A-m(\cup^n_{i=1}c_i))]^M[\frac{c-V_0}{|c-V_0|}+V_0]^{\Sigma}[0]^{\Sigma'}[0]^{\Sigma''}$$

    $\mathcal{H}_2:\Omega\rightarrow\Omega$~(Suspend the box by $V'_0$)
    $$\mathcal{H}_2
    [\rho h_0(A-m(\cup^n_{i=1}c_i))]^M[\frac{c-V_0}{|c-V_0|}+V_0]^{\Sigma}[0]^{\Sigma'}[0]^{\Sigma''}$$$$=[\rho h_0(A-m(\cup^n_{i=1}c_i))]^M[\frac{c-V_0}{|c-V_0|}+V_0]^{\Sigma}[\frac{c-V'_0}{|c-V'_0|}+V'_0]^{\Sigma'}[0]^{\Sigma''}$$

    $\mathcal{H}_3:\Omega\rightarrow\Omega$(Extend the unit vectors: Get the point of intersection)
    $$H_3[\rho h_0(A-m(\cup^n_{i=1}c_i))]^M[\frac{c-V_0}{|c-V_0|}+V_0]^{\Sigma}[\frac{c-V'_0}{|c-V'_0|}+V'_0]^{\Sigma'}[0]^{\Sigma''}$$$$=[\rho h_0(A-m(\cup^n_{i=1}c_i))]^M[\frac{c-V_0}{|c-V_0|}+V_0]^{\Sigma}[\frac{c-V'_0}{|c-V'_0|}+V'_0]^{\Sigma'}[c]^{\Sigma''}$$

    $\Delta:\Omega\rightarrow\Sigma^{+}$
    $$\Delta[\rho h_0(A-m(\cup^n_{i=1}c_i))]^M[\frac{c-V_0}{|c-V_0|}+V_0]^{\Sigma}[\frac{c-V'_0}{|c-V'_0|}+V'_0]^{\Sigma'}[c]^{\Sigma''}=c$$

    The time the system cost from oscillating to stillness can be bounded by a constant. Because of the same reason this method does not break up the lower bound of Turing Machine. But for some other things, we tend to pay more attention to it. Some relevant issues will be discussed in Sec-V.

    \subsubsection{Graph Isomorphism, Graph Spectrum and Oscillators}
    In this part of the section, we shall talk about a complex example in detail. We do not mean to show that the method we designed here is superior to all of the TMs constructed by the people of the same aim. We just want to demonstrate a new style of computation.

    \paragraph{Spectrum of Graph}
    Suppose $X=(V,E)$ is a graph, $A$ is it's adjacent matrix. We say $f_A(\lambda)$ is the characteristic polynomial of $X$, also denoted by $f_X(\lambda)$. $(\lambda_1,\dots,\lambda_n)$, the whole root of $f(\lambda)$, is called the spectrum of graph $X$.

    Actually two different adjacent matrices may represent two isomorphic graphs. If we alter the permutation of the number of the vertices, $A$ will become $P^{-1}AP$, where $P$ is the corresponding permutation matrix. However, the characteristic polynomials of them are the same. Therefore, $f_X(\lambda)$ and the spectrum $\textrm{spec}(X)=(\lambda_1,\dots,\lambda_n)$ are uniquely determined by $X$.

    For the relationship between spectrum and graph, people conjectured that graph can be uniquely determined by spectrum, i.e. suppose
    $$\textrm{spec}(A)=\textrm{spec}(B),$$
    can we conclude that
    $$ A\backsimeq B?$$

    Unfortunately, the different graphs of the same spectrum were found soon.

    Nonetheless, calculating the spectrum is also important. Because we can know a lot of crucial properties, such as the extensionality, rapid mixing time of Markov chains on the graph, by the spectrum of the graph. What's more, when two graph have same spectrum, and spectrum is never repeating, we have a polynomial time algorithm to check whether they are isomorphic.
    \begin{enumerate}
      \item Input graphs~$G_1$，$G_2$, compute their spectrum, denoted by~$\Lambda_1$、$\Lambda_2$.
      \item Compare the spectrums, if $\Lambda_1\neq\Lambda_2$, then return NOT ISOMORPHIC;else, continue;
      \item Check whether the product of the two similar matrices is a permutation matrix, if it is true return ISOMORPHIC, otherwise return NOT ISOMORPHIC;
    \end{enumerate}
    Notation: Here by $\Lambda_1\neq\Lambda_2$ we mean after sorting their eigenvalue, the two series are not identical to each other. And accordingly $G_1$, $G_2$ should also be altered into $\widetilde{G}_1$, $\widetilde{G}_2$. But for convenience, we do not differeciate $G_i$ and $\widetilde{G}_i$.
    \textbf{Proof:}
     \begin{enumerate}
      \item []
      If~$\Lambda_1\neq\Lambda_2$, then~$G_1\ncong G_2$. So we only consider the case in which $\Lambda_1=\Lambda_2=\Lambda$.\\
      i.e.Suppose$$G_1=P\Lambda P^T,~G_2=Q\Lambda Q^T$$
      then we have$$P^TG_1P=\Lambda=Q^TG_2Q$$
      thus
      $$G_1=\left(QP^T\right)^TG_2\left(QP^T\right)$$
      by the preconditon,$\Lambda$~is never repeating, so~$P$，$Q$~is the unique orthganol matrices.\\
      the rest is to show that if~$G_1,G_2$~is isomorphic, then~$QP^T$~is the permutation matrix desired.\\
      In fact, if~$G_1\cong G_2$,then there exists a permutation matrix $S$ s.t.$$G_1=S^TG_2S$$
      Since~$G_2=Q\Lambda Q^T$, the formula above means
      $$G_1=S^TQ\Lambda Q^TS=\left(Q^TS\right)^T\Lambda\left(Q^TS\right)$$
      Because of the uniqueness of~$P$, we can conclude that~$Q^TS=P^T$, and by orthgonality of $Q$, we obtain $$S=QP^T.$$

    \end{enumerate}
    $\square$\\
    \paragraph{Harmonic Oscillator of multi-freedom}
    Suppose $s$ is the number of freedom of the system, $q_{\alpha 0}(\alpha=1,2,\dots,s)$ is the general coordinates when the system is in balance. Without lose of generality, we can always assume that $q_{\alpha 0}$ is just zero, i.e. $q_{\alpha 0}=0(\alpha=1,2,\dots,s)$.

    Because we only talk about little vibration, so we only keep several terms in the Taylor series of the Lagrangians $L$ of the system about $q_{\alpha 0}$.

    The potential energy:
    $$ V=V_0+\sum^s_{\alpha=1}\left(\frac{\partial V}{\partial q_{\alpha}}\right)_0 q_{\alpha}+\sum^s_{\alpha=1}\sum^s_{\beta=1}\frac{1}{2}\left(\frac{\partial^2 V}{\partial q_{\alpha}\partial q_{\beta}}\right)_0 q_{\alpha}q_{\beta}+\cdots.$$
    Note that~$V_0$~can be omitted. Introduce the notation~$k_{\alpha\beta}$，
    $$k_{\alpha\beta}=k_{\beta\alpha}=\left(\frac{\partial^2 V}{\partial q_{\alpha}\partial q_{\beta}}\right)_0,$$
    which is called the strength coefficient. According to the formula~$\left(\frac{\partial V}{\partial q_{\alpha}}\right)_0=0$, the second order of the potential energy could be represented as
    $$ V=\frac{1}{2}\sum^s_{\alpha=1}\sum^s_{\beta=1}k_{\alpha\beta}q_{\alpha}q_{\beta}.$$
    Then assume~$\bm{r}_i=\bm{r}_i(q)$~is not relevant to time, i.e. the obligation is constant, so the kinetic energy is:
    $$ T=\frac{1}{2}\sum^n_{t=1}m_i\dot{\bm{r}}_i\cdot\dot{\bm{r}}_i=\frac{1}{2}\sum^n_{i=1}\sum^s_{\alpha=1}\sum^s_{\beta=1}m_i\frac{\partial \bm{r}_i}{\partial q_{\alpha}}\cdot\frac{\partial \bm{r}_i}{\partial q_{\beta}}\dot{q}_{\alpha}\dot{q}_{\beta}.$$
    Introduce the symbol~$m_{\alpha\beta}$,
    $$ m_{\alpha\beta}=m_{\beta\alpha}=\sum^n_{i=1}m_i\frac{\partial \bm{r}_i}{\partial q_{\alpha}}\cdot\frac{\partial \bm{r}_i}{\partial q_{\beta}},$$
    then the kinetic energy could be represented as
    $$ T=\frac{1}{2}\sum^s_{\alpha=1}\sum^s_{\beta=1}m_{\alpha\beta}\dot{q}_{\alpha}\dot{q}_{\beta}.$$
    Keep the formula above to second order and since~$\dot{q}_{\alpha}\dot{q}_{\beta}$~is second order $m_{\alpha\beta}$ should be expanded to zeroth order. In other words~$m_{\alpha\beta}$ could be looked as constants, we just take the value of them when the system is in balanced point.

    So the Lagrangian could be written as
    $$ L=\frac{1}{2}\sum^s_{\alpha=1}\sum^s_{\beta=1}(m_{\alpha\beta}\dot{q}_{\alpha}\dot{q}_{\beta}-k_{\alpha\beta}q_{\alpha}q_{\beta}).$$
    Thus the Lagrangian equation is
    $$\frac{d}{dt}\frac{\partial}{\partial \dot{q}_{\alpha}}\left(\frac{1}{2}\sum^s_{\beta=1}\sum^s_{\gamma=1}m_{\beta\gamma}\dot{q}_{\beta}\dot{q}_{\gamma}\right)-\frac{\partial}{\partial q_{\alpha}}\left(-\frac{1}{2}\sum^s_{\beta=1}\sum^s_{\gamma=1}k_{\beta\gamma}q_{\beta}q_{\gamma}\right)=0.$$
    i.e.
    $$\frac{d}{dt}\left(\frac{1}{2}\sum^s_{\gamma=1}m_{\alpha\gamma}\dot{q}_{\gamma}+\frac{1}{2}\sum^s_{\beta=1}m_{\beta\alpha}\dot{q}_{\beta}\right)+\left(\frac{1}{2}\sum^s_{\gamma=1}k_{\alpha\gamma}q_{\gamma}+\frac{1}{2}\sum^s_{\beta=1}k_{\beta\alpha}q_{\beta}\right)=0.$$
    therefore
    $$\sum^s_{\beta=1}m_{\alpha\beta}\ddot{q}_{\beta}+\sum^s_{\beta=1}k_{\alpha\beta}q_{\beta}=0\quad (\alpha=1,2,\dots,s).$$
    Let
    $$q_{\beta}=A_{\beta}e^{\lambda t}\quad(\beta=1,2,\dots,s). $$
    Take it into the former formula, we get the linear equations for $A_{\beta}$.
    $$ \sum^s_{\beta=1}(m_{\alpha\beta}\lambda^2+k_{\alpha\beta})A_{\beta}=0\quad (\alpha=1,2,\dots,s).$$
    If the equations have non-trivial solutions, then following conditions should be hold:
    $$\begin{vmatrix}
        m_{11}\lambda^2+k_{11} & m_{12}\lambda^2+k_{12} & \cdots & m_{1s}\lambda^2+k_{1s}  \\
        m_{21}\lambda^2+k_{21} & m_{22}\lambda^2+k_{22} & \cdots & m_{2s}\lambda^2+k_{2s} \\
        \vdots & \vdots &  & \vdots \\
        m_{s1}\lambda^2+k_{s1} & m_{s2}\lambda^2+k_{s2} & \cdots & m_{ss}\lambda^2+k_{ss} \\
      \end{vmatrix}=0
     $$
     This is the equations of times $s$ of $\lambda^2$, and we can get $s$ $\lambda^2$, denoted by

     $$\lambda^2_l\quad(l=1,2,\cdots,s).$$
     \paragraph{The characteristic oscillators for a Graph}
     Making use of the conclusions above, we construct a specific oscillators for any given connected graph.

     Denote the vertices of graph by numbers $1\sim n$, according to any order. The mass of a vertex is set 1g.
     Connect the vertex $1$~and~$n$ to ends by strings whose $k$ is zero by that direction.
     For the rest, we connect them according to the adjacent matrix, i.e. if
     ~$A_{ij}=1(\textrm{Note that }A_{ij}=A_{ji})$, connect vertex~$i$ and $j$ by a string whose $k=1$.
     Let's study the motion of the system:
     First, if two vertices is not connected by string, we have $k_{\alpha\beta}=k_{\beta\alpha}=0$. Second, the vibration is little, so string is
     not an obligation. And we take the general coordinates as the usual displacement vectors, so $m_{\alpha\beta}=m_{\beta\alpha}=\delta_{\alpha\beta}$, where $\delta_{\alpha\beta}$ is the well known Kronecker notation.

     At last, we obtain the determinant as follows, which is the characteristic polynomial of our system.
      $$\begin{vmatrix}
        \lambda^2+d_1 & -A_{12} & \cdots & -A_{1s}  \\
        -A_{21} & \lambda^2+d_2 & \cdots & -A_{2s} \\
        \vdots & \vdots &  & \vdots \\
        -A_{s1} & -A_{s2} & \cdots & \lambda^2+d_s \\
      \end{vmatrix}=0
     $$
     It has been proofed that $\lambda^2<0$. So let~$-\Lambda=\lambda^2$~,we can see that the determinant above actually compute the spectrum of $A'$ which is converted from~$A$~by adding multi-loops(the number of degrees). If a vertex is in the characteristic position, it will take part in the vibrations of all frequencies, if no one is in the characteristic position, then they just vibrate with respective frequency. In both cases, we'll measure the frequency and differentiate them by means of FFT, so as to get the spectrum of $A'$.

     Apparently, adding multi-loops is not harmful to the decision of whether $A$ and $B$ are isomorphic, for if $A'\neq B'$, then $A\not\backsimeq B$
     If $A\backsimeq B$,
     then ~$A'\backsimeq B'$, which will also be checked by the oscillating system.

     \paragraph{Comments}
     Suppose now the problem we want to solve is: try to find the $n_0-$th($n_0$~is a constant) value of such matrices. It is easy to see that the period of the system satisfies $$\frac{1}{\sqrt{n}f_0}\leq T\leq \frac{\sqrt{n}}{f_0}(\text{or }\frac{1}{\sqrt{n}}f_0\leq f\leq \sqrt{n}f_0),$$ where $f_0$ is the eigenfrequency of a single string, considering the minimum case occurs when the corresponding graph is totally parallel connected(two vertices with $n-$multiple edges between them), while the maximum case occurs when it is just a chain.

     The total steps of sampling should be $$N=2BL=2\left(\sqrt{n}f_0-\frac{f_0}{\sqrt{n}}\right)\left(c_{n_0}\frac{\sqrt{n}}{f_0}\right)$$

     where the constant $c_{n_0}$ is related to the required precise $n_0$. So we have: $O(N)=O(n)$ and the corresponding complexity for fast fourier transform should be $O(n\ln n)$.

      So we can say that the time complexity of this method should be $$O(n^2)+O\left(\frac{\sqrt{n}}{f_0}\right)+O(n\ln n)+O(n)=O(n^2)$$
     where the left $O(n^2)$ is the cost of constructing the system and $O(n\ln n)$ is the complexity of FFT. On the other hand, if we use the well-known algorithm called QR-method to get the answer, it will cost such Turing machine $O(n^3)$ steps. However, we know little about wether our method is superior to any most efficient Turing machines.

     However, such analysis may not be enough. We cannot obtain a reliable result unless some specifications for the materials are considered. The procedure of constructing the system according to an arbitrary graph is actually quite tricky. For this system need $n$ oscillators of the same mass and different length. To achieve this, we have to assume that there exists a kind of `ideal' material which, at least for a large enough range, can be stretched to a desired length easily(in $O(n)$ time and totally $O(n^2)$) and, at the same time, keep rigid.

    \subsubsection{Steiner Tree Problem}
    Steiner Tree Problem is a problem in combinatorics. The general version of Steiner Tree Problem is NP-complete, which implies that this problem is unlikely be solved in polynomial time.

    This problem is similar to the Minimal Spanning Tree Problem in metric space. The difference is that Steiner Tree Problem allow people to add new points $v'(v'\not\in V)$ and new edges $e'(e'\not\in E)$ into the original graph $G$, if necessary. When $|G|=3$, the new point (in this case, at most one point is needed)is called Fermat point.

    At a time, some people became to believe that the experiments of soup membrane can be used to solve the Steiner Tree Problem. In fact, when $|G|$ is small, say, less than 5, this method really works. However, when the number of vertices is 10 or more, this experiment just cannot give the right answer. One can attribute the failure to different reasons and derive various explanations. Of all these potential explanations, the one which states that `it is just the errors during the experiment cause the failure' made many people conjecture faithfully that classical mechanics can be used to solve NP-complete Problems in polynomial time(So they try to proof P$=$NP).

    In fact, the foundation of the experiment is the well-known property that the membrane will stay at a stationary state, where the surface it produces will be just the minimal surface. Unfortunately, this theory has nothing to do with the fact that the membrane can arrive at the stationary state \emph{fast}. What's more, no one can proof the soundness of such property under the framework of classical mechanics.

    \subsubsection{DNA Computation}
    In 1994, Adleman used a probabilistic DNA algorithm to solve HP problem (Hamilton Path Problem). HP problem is NP-complete, which implies it is difficult to find a polynomial algorithm to solve it\cite{adl}.

    In order to understand Adleman's method, the following knowledge seems necessary.
    \begin{enumerate}
      \item [(1)]DNA contains chains consisted by four types of nucleotides, denoted by A, C, G and T.
      \item [(2)]These nucleotides forms complementary couples, i.e. A and T are complementary, C and G are complementary. If the corresponding positions of two DNA chains are complementary, they will patch up as the twin-helix structure.
      \item [(3)]PCR, which proposed by Kary Mullis, is method to reproduce the specific chain we need.
      \item [(4)]There is a machine called 'sequencer' which can be used to read out the series of a DNA chain.
    \end{enumerate}
    Adleman's Algorithm contains five procedures(Suppose $|G|=n$):
    \begin{enumerate}
      \item [(1)]Randomly produce the paths in the Graph, encoded by DNA chains.
      \item [(2)]Keep only those paths which begin with $v_{in}$ and end with $v_{out}$.
      \item [(3)]Keep only the paths whose length is $n$
      \item [(4)]Keep only those paths which enter all vertices in G at least once.
      \item [(5)]If any paths remain, return 'True', else return 'False'.

    \end{enumerate}
     Note that the first step of Adleman's Algorithm, which is usually thought to be work as an initialization operator $\nabla$, is not polynomial with respect to the resource mass $\mathfrak{M}$ and space $\mathfrak{S}$ at least. Considering asymptotically we can only sequentially get the mass the algorithm need, so actually $O(n!)$ mass can cause $O(n!)$ time $\mathfrak{T}$. As a matter of fact the other steps of this algorithm, which require exponentially molecules fully blend by polynomially increasing contacting facades, also cost a lot of resource $\mathfrak{T}$.

     It is not very hard to appreciate the conclusion that we can obtain great power of computation suppose we are provided with corresponding quantity of mass, and do not take the cost of preparing such equipment at all. For one thing, let's consider the following ideal model.

     Suppose we have enough universal Turing Machines, each of them are denoted by their footnotes. What's more, by some altering in the definition, these UTMs have the ability to transmit their results to others. And the condition of two UTMs $U_i, U_j(i\neq j)$ could communicate to each other is that they are adjacent to each other, denoted by $Adj(U_i,U_j)$.

     So the computational model constructed following, called `Turing Tree', can exponentially speed up the computation of any NP-complete problems.

     \begin{defin}(Turing Tree)
     Suppose we have infinite many UTMs, each of them denoted by unique footnotes, and
     $$Adj(U_i,U_j)\Leftrightarrow j=2i+1\vee i=2j+1\vee j=2i+2\vee i=2j+2,$$ then we call this Turing Tree.
     \end{defin}
     It is easy to see that the following relation holds:
     $$\begin{array}{c}
                  Adj(U_0,U_1),Adj(U_0,U_2) \\
                  Adj(U_1,U_3),Adj(U_1,U_4),Adj(U_2,U_5),Adj(U_2,U_6)\\
                  \dots\dots\dots
                \end{array}
      $$
     For example, a TSP problem can be solved as following:
     \begin{enumerate}
        \item [a]The Observer input the weighted complete graph $G$ to the $U_0$, $U_0$ decode 0 to a permutation and compute the sum of the weight, and then transmit $G$ and flag $F=0$ to $U_1,U_2$.
        \item [b]For index $i$，After $U_i$ get $F=0$ and $G$, it check whether $i<\lceil\log_2 n!\rceil$,if the answer if 'yes' then decode $i$ to a permutation and get the sum, and transmit $G$ and $F=0$ to $U_{2i+1}$ and $U_{2i+2}$;else check whether $i=\lceil\log_2 n!\rceil$, if it is true, decode $i$ to a permutation and get the sum, then submit the weight sum to the $U_{\lceil i/2\rceil-1}$. Else, do nothing.
        \item [c]For index $i$, after $U_i$ get $F=1$ and two sum(come from $U_{2i+1},U_{2i+2}$), if it's index is not zero, then submit $\min\{S_i,S_{2i+1},S_{2i+2}\}$ and $F=1$ to $U_{\lceil i/2\rceil-1}$. Else return $\min\{S_0,S_1,S_2\}$ and write it on to the tape.
      \end{enumerate}

      It is easy to check that the subprocedure of the algorithm which is used to decode a natural number to a permutation is polynomial. So the cost of time the Turing Tree consumed should be $O(2\log_2n!)\leq O(2n\log_2n)$ (Including once sharing the task and once championship for the minimum). So it is the time to answer how can we `easily' construct a big enough Turing Tree.

    \subsection{Preliminary Discussion of the classic theory of Computation}
    \subsubsection{Turing computable is physical computable}
    The topic about the existence of a theoretical physical system which can provide an implementation of universal Turing Machine has been studied by many scholars. In addition to the current implementation of computers, scholars have constructed many other wonderful designs on various axiom systems of physics(e.g. Classical Mechanics, Quantum Mechanics).

    Of course, the results above only imply that it is the ideal mathematic model for a family of physical phenomenons can be look as equivalent to UTM in terms of computability. After all, we cannot know for sure that some theory of physics is completely correct. Because of this, when we talk about the ability of computation for a certain family of physical system, we always assume either of the two preconditions:
    \begin{itemize}
      \item The ideal mathematic model of some branch of physics is believed to be absolutely right.
      \item At least in a very large scale, the theory works.
    \end{itemize}

    \subsubsection{PLATO Machine}
    For several decades after the Church-Turing Thesis was proposed, people failed to find a counter-example. This kind of counter-example, if they really exist, should satisfies the property that most people think they can be effectively computed in principle, and no Turing machine can compute them.

    However, many physicists tend to make efforts in another direction, that is, they want to find a family of processes in nature, whose functional expression may not be intuitively computable, nor Turing Computable, but actually it can be used to `compute' a nonrecursive function by measurement.

    Suppose the problem we attempt to deal with now may cost infinite many steps for some computational model(e.g. Turing Machine), does it necessarily mean that we have to wait infinitely long time to get the results? This is not always the case, PLATO Machine, which was proposed by H.Weyl\cite{pit}, is just a counter-example. Though it is named after Plato, the designer's main inspiration comes from one of Zeno's Paradoxes.

    Specifically, PLATO Machines use $(1/2)^{n}$ seconds to execute the $n-$th step. For instance, suppose the decision problem we want to solve is $\exists nP(n)$, where $P$ is a predicate and $P(x)$ is used to describe some properties of $x$. Then PLATO machine $\mathbb{P}$ will check whether $P(1)=1$ holds in $1/2$ seconds, and check whether $P(2)=1$ in $1/4$ seconds,$\dots$, and check whether $P(n)$ holds in $2^{-n}$ seconds, and so on. It is easy to conclude that if $\mathbb{P}$ find an answer, it will return the answer in one second, otherwise it will return false after a second. Considering the sum of geometric series, the proof is trivial. So the upperbound of the time for $\mathbb{P}$ to solve any question is $$T=\left(\frac{1}{2}\right)^1+\left(\frac{1}{2}\right)^2+\cdots+\left(\frac{1}{2}\right)^n+\cdots=\frac{\frac{1}{2}}{1-\frac{1}{2}}=1s$$
    Apparently, if $\mathbb{P}$ does exist, its power is extraordinarily great, for it can even solve Turing's Halting Problem in one second.

    So far we have seen two idea to implement the PLATO machine $\mathbb{P}$. However, unfortunately, neither of them are successful. The first one is to construct the machine according to the definitions of H.Weyl. Apparently, it is difficult, for people do not believe that time is infinitely divisible. The second one is to make use of the theory of general relativity. However, the computing system will also exhaust the resource of the universe which make the observer cannot get the answer.

    \subsubsection{Recursive function whose derivative is not recursive}
    April 1970, J.Myhill published his astonishing result\cite{myh}: There exists a recursive function, whose derivative is not recursive. In order to understand the principles of the construction, knowing the following fact about the recursive functions(whose domain is $\mathbb{R}$) should be helpful.
    \begin{theorem}
    Suppose $f$ is a real-valued function, $\{f_n\}$ is a series of recursive functions, if there exists a recursive function $e:\mathbb{N}\rightarrow \mathbb{N}$ s.t. $\forall x\in I$，$k\geq e(n)$~有~$|f_k(x)-f(x)|\leq \frac{1}{2^n}$, then $f$ is recursively computable.
    \end{theorem}

    J.Myhill's idea is to build a non-trivial structure(slope or bump) in the neighborhood of $2^{-n}$ in interval $[0,1]$, where $n\in\mathscr{A}$, and $\mathscr{A}$ is a recursively enumerable, nonrecursive set. Otherwise $f(x)=0$. However, in order to make the function computable, the scale of the structure should shrink as the $n$ is enumerated recursively, or rather, should be smaller than the bound in the theorem above. As a result, the derivative of the function is intuitively hard to compute, and on the other hand we can proof that it is not recursive, because if we could compute it we can use the result to decide whether $\lceil x\rceil$ is an element of $\mathscr{A}$ generally, contradicting the nonrecursiveness of $\mathscr{A}$.

    Specifically, suppose$$\theta(x)\equiv\left\{
                                    \begin{array}{ll}
                                      x(x^2-1)^2, & \hbox{if~$-1\leq x\leq 1$;} \\
                                      0, & \hbox{if~$|x|>1$.}
                                    \end{array}
                                  \right.
    $$

    It is easy to verify that $\theta(-1)=\theta(0)=\theta(1)=0$，$\theta'(-1)=\theta'(1)=0$，$\theta'(0)=1$，and~$\theta_{min}=\theta(-1/\sqrt{5})\equiv-\lambda$，$ \theta_{max}=\theta(+1\sqrt{5})\equiv+\lambda$. We call~$\theta$~a $bump$ of length 2 and height $\lambda$.
    Then the function $\theta_{\alpha\beta}(x)\equiv(\beta/\lambda)\theta(x/\alpha)$ satisfies the following conditions:
    $$\theta_{\alpha\beta}(-\alpha)=\theta_{\alpha\beta}(0)=\theta_{\alpha\beta}(\alpha)=0,\quad \theta'_{\alpha\beta}(-\alpha)=\theta'_{\alpha\beta}(\alpha)=0,\quad \theta'_{\alpha\beta}(0)=\theta/\lambda\alpha,$$
$$-\beta\leq\theta_{\alpha\beta}(x)\leq\beta\quad(\alpha\leq x\leq \alpha.)$$

    For each~$n\in\mathscr{A}$，we shall construct a$bump$: $\theta_{\alpha_n\beta_n}$ at $2^{-n}$ i.e.

    if~$n\in\mathscr{A}$, ~$\delta\in[-\alpha_n,+\alpha_n]$, $f(2^{-n}+\delta)\equiv\theta_{\alpha_n\beta_n}(\delta)$, otherwise~$f(x)\equiv0$.
    To make~$f$~well-defined, parameters~$\alpha_n,\beta_n,n\in\mathscr{A}$~is defined as $$\alpha\equiv2^{-k-2n-2},\quad \beta_n\equiv2^{-k-n-2},$$
    where $n=h(k)$~and $h$~is a function enumerating~$\mathscr{A}$~without repetitions(It is easy to proof that if there exists a recursive function enumerating $\mathscr{A}$, then there exists such function with no repetitions).

    For physicists, does J.Myhill's results imply that if an object move under the condition that the displacement and the time satisfies the following relations
    $$\textbf{\emph{r}}(t)=f(t)=\left\{
                                                             \begin{array}{ll}
                                                               \theta_{\alpha_n\beta_n}(\delta)(n\in\mathscr{A}), & \hbox{if~$t=2^{-n}+\delta$,\quad $\delta\in[\-\alpha_n,+\alpha_n]$;} \\
                                                               0, & \hbox{o.w.}
                                                             \end{array}
                                                           \right.
$$
The speed ~$\textbf{\emph{v}}\equiv \textbf{\emph{r}}'(t)$~will be a physical quantity which is not computable?
    \subsubsection{Physical States which is not computable}
    Pour-El et al published their results in 1997: for a differential equation, one can design a specific initial state to make the solution after $t$ ($t$ could be take some computable value)seconds is nowhere computable\cite{pou}.

    Consider the IVP of the following wave equation:
    $$\left\{
        \begin{array}{ll}
          \frac{\partial^2u}{\partial t^2}=\frac{\partial^2 x}{\partial x^2}+\frac{\partial^2 y}{\partial y^2}+\frac{\partial^2 z}{\partial z^2}, & \hbox{} \\
          u(x,y,z,0)=f(x,y,z), \frac{\partial u}{\partial t}(x,y,z,0)=0& \hbox{.}
        \end{array}
      \right.
    $$
    where $(x,y,z)\in \mathbb{R}^3,\quad t\in[~0,+\infty~)$
    for all $f\in \mathscr{C}^1$ this IVP has a form of solution known as Kirchhoff's formula:
    $$u(\overrightarrow{x},t)=\iint_{S^2}[f(\overrightarrow{x}+tn)+t\nabla f(\overrightarrow{x}+t\overrightarrow{n})\cdot \overrightarrow{n}]d\sigma(\overrightarrow{n})$$
    The conclusion Pour-El get is the following theorem:
    \begin{theorem}
    For all compact set $D\subset\mathbb{R}^3\times[~0,\infty~)$, there exists a computable function $f(x,y,z)\in\mathscr{C}^1$, s.t. the corresponding solution $u(\overrightarrow{x},t)$ is not computable in the neighborhood of any point in $D$.
    \end{theorem}
    Pour-El et al construct the initial value through the uncomputable real number $\sum^{\infty}_{i=0}\frac{1}{2^{a(i)}},\quad a(i)\in \mathscr{A}$.

    Apparently, one can conclude that in this wave equation, the initial state is computable but the state $u(0,0,0,1)$ is a state which can not be compute.

    For us, can we safely conclude that $$\{\text{Turing Computable}\}\subset\{\text{Physical Computable}\}$$
    but $$\{\text{Turing Computable}\}\not\supset\{\text{Physical Computable}\}?$$

    \subsubsection{A few Comments}
    In the above scenario, the use of (actual) infinity is their common theme. They ask the system to run for infinite steps or just encode the solutions into real numbers. It is easy to find out that adding either of these two assumes into a physical system will make the original system extraordinarily powerful.

    For example, we can throw a particle onto a plane $[0,1]\times[0,1]$ at random(obey the uniform distribution), then we can proof that with high probability, the $x$-coordinate(or $y$-coordinate) of the center of the particle will indicate a non-recursive real number. In fact, in cell $[0,1]\times[0,1]$, the Lebesgue measurement for the recursive real numbers is 0, while the rest is 1, i.e.
    $$m([0,1]\times[0,1]\cap\mathbb{R}_{r})=0, m([0,1]\times[0,1]\cap\mathbb{R}^c_{r})=1$$
    This is geometric probability and consider the uniform distribution, the probability of the either event of the two are just their measurement. Therefore we can look the $x-$coordinate as a function with respect to the digits. According to Beggs et al, a theoretical machine called SME may help us to get the value of the position coordinates\cite{beggs2}\cite{beggs3}\cite{beggs4}.

    However, does the strict plane really exist in the physical world? We just do not know.

    We propose the some levels for $f$ which is not computable. Suppose $o$ is an operator, $\hat{o}$ is a physical implementation of $o$ and operators $\Delta$ , $\nabla$ always exist.
    For $f\in F$ where all elements in $F$ are non-recursive functions, we have the level of existence as follows.
    \begin{itemize}
      \item \textbf{Existence-I} \\
          $\exists p(\Delta\circ p\circ\nabla\doteq f)$

      \item \textbf{Existence-I*}\\
          $\exists \hat{p}\exists\hat{\nabla}\exists F((f\in F)\wedge(\Pr\{\Delta\circ\hat{p}\circ\hat{\nabla}\doteq f'|f'\in F\}>0))$
      \item \textbf{Existence-II}\\
          $\exists\hat{\Delta}\exists p(\hat{\Delta}\circ p\circ\nabla\doteq f)$

      \item \textbf{Existence-II*}\\
          $ \exists \hat{p}\exists\hat{\nabla}(\Delta\circ \hat{p}\circ\hat{\nabla}\doteq f)$

      \item \textbf{Existence-III}\\$ \exists \hat{p}\exists\hat{\Delta}\exists\hat{\nabla}(\hat{\Delta}\circ \hat{p}\circ\hat{\nabla}\doteq f)$

    \end{itemize}
    According to the levels we proposed above, assume the space is continuous, we can find out that $\mathbb{P}\in\text{Existence-II}$, J.Myhill's function $f\in\text{Existence-I}$, Pour-El's construction $\phi\in\text{Existence-II}$, our example $x\in\text{Existence-I*}$. Apparently, we wish to get the examples in $\text{Existence-III}$.

    \section{Physical Resource Complexity for Quantum Computation}
    \subsection{Physical Resource Complexity for Quantum Computation}
    For general quantum computation, we only need to explain the definition of the physical state set and the required evolution operators. More over, we only talk about Monte Carlo styled quantum algorithms.

    According to von Neumann's four postulates for quantum mechanics, we require that the state of any representation should be vectors in Hilbert space,i.e.$$\Omega\subset \mathbb{H}$$ and the evolution operators should be unitary, i.e. $$\mathcal{H}\in U(n)$$ Without loss of generality, we can assume that the measurement operators is projection operators(POVM could be substituted by projection operators through adding more auxiliary qubits)

    \subsubsection{The RCEF for Quantum Computation}
    The resource cost by a computation is $$\begin{array}{ccc}
    \mathfrak{R}_{\mathscr{P}}(\Delta\circ\mathcal{H}\circ\nabla)&= & (\mathfrak{T}(\Delta\circ\mathcal{H}\circ\nabla), \\
                                                                                                           &  & \mathfrak{S}(\Delta\circ\mathcal{H}\circ\nabla), \\
                                                                                                           &  & \mathfrak{E}(\Delta\circ\mathcal{H}\circ\nabla), \\
                                                                                                          &   & \mathfrak{G}(\Delta\circ\mathcal{H}\circ\nabla))
                                                                                                         \end{array}
    $$
    Our definition here is special a case of the one in Sec-III. Suppose our discussion is restricted to QCM, i.e. we have finite kinds of universal quantum operators, then the number of gates used and the depth of the whole circuit will be the main parameter which should be took into account. It is easy to find out that this definition is similar to that of quantum circuit model. One of the difference between them is that we will also take the cost of design(usually this costs time) of a new circuit into account. Though in most cases, this will not cause great difference from the result given by QCM, however, we don't think we can safely ignore the potential exceptions just because it is usually easy to expand the scale of some circuits.

    So far, people always assume that qubit is relatively easy to prepared. At least in the asymptotic sense, no matter how difficult to prepared a quantum bit, the cost should be bounded by a constant. We will also do this.

    \subsubsection{Deutsch-Josza Algorithm}
    Deutsch-Josza algorithm is one of the most successful algorithms in the early years. The corresponding problem of the algorithm is: consider two sets of functions:
    \begin{enumerate}
      \item [A:]$\bigg\{\varphi|\varphi:\{0,\dots,2^n-1\}\rightarrow\{0,1\}\textrm{,~且~}\forall x(\varphi(x)=0)\bigg\}$
      \item [B:]$\bigg\{\varphi|\varphi:\{0,\dots,2^n-1\}\rightarrow\{0,1\}\textrm{,~且~}\Big|\{x|\varphi(x)=0\}\Big|=\Big|\{x|\varphi(x)=1\}\Big|\bigg\}$
    \end{enumerate}
    Apparently we have $A\cap B=\emptyset$, now suppose~$f\in A\cup B$ and there is an oracle to compute~$f$. We are required to decide whether $f\in A$ or not. It is no doubt that people wish to reduce the times of query the oracle as much as possible.

    Note that the cost of implement the oracle is not taken into account, because we assume we have implemented it.

    The algorithm needs a trivial input ~$\psi_0=|0\rangle^{\otimes n}|1\rangle$, and used the gate~$H^{\otimes n}\otimes H$~onto~the state$\psi_0$~a nd get~$\psi_1$, i.e.
    $$\psi_1=\Big(H^{\otimes n}\otimes H\Big)\left(|0\rangle^{\otimes n}|1\rangle\right)$$
    Note that~$H=\frac{1}{\sqrt{2}}\Big((|0\rangle+|1\rangle)\langle 0|+(|0\rangle-|1\rangle)\langle 1|\Big)$. By induction we have
    $$H^{\otimes n}=\frac{1}{\sqrt{2^n}}\sum_{x,y}(-1)^{x\cdot y}|x\rangle\langle y|$$
    where i.e.~~$x\cdot y\equiv\bigoplus\limits_i x_i\wedge y_i$.
    So we get:
    $$\begin{array}{lll}
        \psi_1&= & \Big(H^{\otimes n}\otimes H\Big)\left(|0\rangle^{\otimes n}|1\rangle\right) \\
              &= & \frac{1}{\sqrt{2^n}}\bigg(\sum\limits_{x,y}(-1)^{x\cdot y}|x\rangle\langle y|\bigg)|0\rangle^{\otimes n}\left[\frac{|0\rangle-|1\rangle}{\sqrt{2}}\right] \text{(by orthogonality)}\\
              &= & \frac{1}{\sqrt{2^n}}\sum\limits_x(-1)^0|x\rangle\left[\frac{|0\rangle-|1\rangle}{\sqrt{2}}\right] \\
              &= & \frac{1}{\sqrt{2^n}}\sum\limits_{x=0}^{2^n-1}|x\rangle\left[\frac{|0\rangle-|1\rangle}{\sqrt{2}}\right]  \\
      \end{array}
    $$
    Now use the oracle~$U_f:|x,y\rangle\rightarrow|x,y\oplus f(x)\rangle$ onto the state $\psi_1$ to get $\psi_2$
    $$\begin{array}{lll}
        \psi_2&= & U_f  \Bigg(\frac{1}{\sqrt{2^n}}\sum\limits_{x=0}^{2^n-1}|x\rangle\left[\frac{|0\rangle-|1\rangle}{\sqrt{2}}\right]\Bigg)\\
              &= & \frac{1}{\sqrt{2^n}}\sum\limits_{x=0}^{2^n-1}|x\rangle\bigg(f(x)\oplus\frac{|0\rangle-|1\rangle}{\sqrt{2}}\bigg)\\
               &  &  \\
              &= & \frac{1}{\sqrt{2^n}}\sum\limits_{x=0}^{2^n-1}(-1)^{f(x)}|x\rangle\left[\frac{|0\rangle-|1\rangle}{\sqrt{2}}\right]  \\
      \end{array}
    $$
    At last we use $H^{\otimes n}\otimes I$ onto $\psi_2$to get$\psi_3$:
    $$\begin{array}{lll}
   \psi_3 & = & \Big(H^{\otimes n}\otimes I\Big)\Bigg(\frac{1}{\sqrt{2^n}}\sum\limits_{x=0}^{2^n-1}(-1)^{f(x)}|x\rangle\left[\frac{|0\rangle-|1\rangle}{\sqrt{2}}\right]\Bigg) \\
          & = &  \sum\limits_z\sum\limits_x\frac{(-1)^{x\cdot z+f(x)}|z\rangle}{2^n}\left[\frac{|0\rangle-|1\rangle}{\sqrt{2}}\right]\\
      \end{array}
    $$
    The observer is supposed to check the first $n$ qubits, note that the amplitude of $|0\rangle^{\otimes n}$ is $\sum_x(-1)^{f(x)}/2^n$.
    If $f\in A$, $f(x)$ is constant and the amplitude of $|0\rangle^{\otimes n}$ is $+1$ or $-1$. So the amplitude of another cases should be zero and the observer will get $|0\rangle^{\otimes n}$. On the other hand, if $f\in B$, the amplitude of $|0\rangle^{\otimes n}$ will be zero. So the observer will always get a non-zero vector.

  In our opinion, the procedure could be written as follows.
  $$\begin{array}{lll}
      \nabla&\equiv &\text{Initialize the state} |0\rangle^{\otimes n}\otimes |1\rangle \\
            &       &\text{Generate the whole circuit}\\
         \mathcal{H}_1&\equiv & H^{\otimes n}\otimes H\\
         \mathcal{H}_2&\equiv & U_f \\
         \mathcal{H}_3&\equiv & H^{\otimes n}\otimes I \\
   \Delta&\equiv & \sum\limits_i|P_i\rangle\langle P_i|
    \end{array}
  $$
  Let $\mathcal{H}=\mathcal{H}_3\circ \mathcal{H}_2\circ \mathcal{H}_1$\\
  $\left(\Delta\circ \mathcal{H}\circ\nabla\right)(|x\rangle)=\Big(\Delta\circ \mathcal{H}_3\circ \mathcal{H}_2\circ \mathcal{H}_1\circ\nabla\Big)(|x\rangle)=P(f\in B)$

  Though Deutsch-Jozsa Algorithm is great, someone still think it is not very useful. In addition to the fact that the problem they studied is not very important, there does exist an efficient classical probabilistic algorithm to solve the problem with high probability.

    \subsubsection{Shor's Algorithm}
    Shor's Algorithms for prime factorization and discrete logarithms\cite{shor1,shor2} is so far the most exciting quantum algorithms. The appearance of Shor's Algorithms is the greatest challenge to strong Church Turing Thesis.

    Shor's Algorithms depends on a technique of so called "quantum Fourier Transform". But of course QFT is not enough. Shor's Algorithm is totally non-trivial and marvelous, and few people can produce any algorithms like that easily.

    In order to understand Shor's Algorithm, it may be enough to gain a clear idea of quantum ordering algorithm. This is the only subprogram in the Shor's Algorithm which has to be implemented by quantum computers so far, and it is really the most important subprogram.

    First, note that$$\sum^{r-1}_{s=0}\exp(-2\pi isk/r)=r\delta_{k0}$$

    and define $|u_s\rangle$ as follows
    $$|u_s\rangle\triangleq\frac{1}{\sqrt{r}}\sum^{r-1}_{k'=0}e^{-2\pi
    isk'/r}\Big|x^{k'}\textrm{mod}N\Big\rangle$$

    According to the fact above, we can get
    $$\frac{1}{\sqrt{r}}\sum^{r-1}_{s=0}e^{2\pi isk/r}|u_s\rangle=\Big|x^k\text{mod } N\Big\rangle$$
    In fact
    \begin{align*}
    \frac{1}{\sqrt{r}}\sum^{r-1}_{s=0}e^{2\pi isk/r}|u_s\rangle
    =&\frac{1}{\sqrt{r}}\sum^{r-1}_{s=0}\left(e^{2\pi
    isk/r}\frac{1}{\sqrt{r}}\sum^{r-1}_{k'=0}e^{-2\pi
    isk'/r}\Big|x^{k'}\textrm{mod}N\Big\rangle\right)\\
    =&\frac{1}{r}\sum^{r-1}_{s=0}\left(e^{2\pi
    isk/r}\sum^{r-1}_{k'=0}e^{-2\pi
    isk'/r}\Big|x^{k'}\textrm{mod}N\Big\rangle\right)\\
    =&\frac{1}{r}\sum^{r-1}_{k'=0}\sum^{r-1}_{s=0}\textrm{exp}\left({\frac{2\pi is(k-k')}{r}}\right)\Big|x^{k'}\textrm{mod}N\Big\rangle\\
    =&\frac{1}{r}\sum^{r-1}_{k'=0}r\delta_{kk'}\Big|x^{k'}\textrm{mod}N\Big\rangle\\
    =&\Big|x^k\text{mod } N\Big\rangle
   \end{align*}
   In particular, when $k=0$, we have$$\frac{1}{\sqrt{r}}\sum^{r-1}_{s=0}|u_s\rangle=|1\rangle^{\otimes L}$$
   where $L\equiv\lceil \log (N)\rceil$.

   Suppose~$U_{x,N}$ satisfies~$U_{x,N}|y\rangle\triangleq|xy(\text{mod}N)\rangle$. Considering~$\mathbb{Z}^*_N$ and the fact that the permutation on orthnormal basis can be represented as a unitary operator, one can know for sure that $U_{x,N}$ is unitary.
   What's more~$u_s$~is a eigenvector of~$U_{x,N}$~, the corresponding eigenvalue is~$e^{\frac{2\pi is}{r}}$~since
   $$U_{x,N}|u_s\rangle=\frac{1}{\sqrt{r}}\sum^{r-1}_{k=0}e^{\frac{-2\pi isk}{r}}\Big|x^{k+1}\text{mod}N\Big\rangle=e^{\frac{2\pi is}{r}}|u_s\rangle$$

   Reverse the results above, we get the first half of the quantum ordering Algorithm, which complete the following task:

   $$|1\rangle^{\otimes L}=\frac{1}{\sqrt{r}}\sum^{r-1}_{s=0}|u_s\rangle\xrightarrow[\text{Modular exponentiation}]{U_{x,N}^{z_t2^{t-1}}\cdots U_{x,N}^{z_12^0}}\boxed{\frac{1}{\sqrt{r}}\sum^{r-1}_{s=0}e^{2\pi isk/r}|u_s\rangle}=\Big|x^k\text{mod } N\Big\rangle$$
   where the state $\frac{1}{\sqrt{r}}\sum^{r-1}_{s=0}e^{2\pi isk/r}|u_s\rangle$ is the one we desire.
   Apparently the eigenvalue contains the information of $r$. So as to extract the information, we need a sub-progress named "quantum phase estimation" which based on inverse quantum fourier transformation. One can verify that if $t$ is large enough, such like $t=2L+1+\lceil\log\left(2+\frac{1}{2\varepsilon}\right)\rceil$, for each $s\in\{0,\dots,r-1\}$, we will obtain the estimation of $\varphi\approx s/r$ accurate to $2L+1$ bits with probability at least $(1-\varepsilon)/r$. Through the continued fractions algorithm, we will get $r$ with high probability(According to PNT).

   In our opinion, the procedure above could be written as:
    $$\begin{array}{lll}
      \nabla&\equiv &\text{Initialize the state} |0\rangle^{\otimes t}\otimes |1\rangle^{\otimes L} \\
            &       &\text{Generate the whole circuit}\\
         \mathcal{H}_1&\equiv & H^{\otimes n}\otimes H\\
         \mathcal{H}_2&\equiv & CU_{x,N} \\
         \mathcal{H}_3&\equiv & FT^{\dag}\otimes I^{\otimes L}\\
         \mathcal{H}_4&\equiv & CF\otimes I^{\otimes L}\\
   \Delta&\equiv & \sum\limits_i|P_i\rangle\langle P_i|
    \end{array}
  $$
    It is easy to check that except the $\mathcal{H}_3$, all operators cost polynomial time with respect to $\log N$.
    The complexity of operator modular exponentiation and continued fraction are both $O(L^3)$, which are two most time-consuming subprocedure of the whole algorithm except the $\mathcal{H}_3$(inverse quantum fourier transform).

    Note that $\mathcal{H}_3$ is indeed not an operator which could be implemented by polynomial universal gates.
     Consider a family of gates used in $\mathcal{H}_3$ which is usually noted by $R_k$($k\in\{2,\dots,L\}$)
     $$R_k=\begin{pmatrix}
        1 & 0 \\
        0 & e^{2\pi i/2^k} \\
      \end{pmatrix}
$$
In other words, the original Shor's Algorithm is not a algorithm with super-polynomial acceleration. In order to overcome this, Coppersmith created a new algorithm called the AFFT(Approximate Fast Fourier transform) \cite{cop} which can substitute for the procedure QFT.

    \subsubsection{Grover Algorithm}
    Quantum Search Algorithm\cite{gro}, also known as Grover's Algorithm, is another quite successful quantum algorithm.
    Though this algorithm is not faster than the fastest classical search algorithms super-polynomially, one can proof it is the fastest one considering quantum mechanics. Therefore, the complexity of the algorithm is the complexity of the problem it deals with.

    The crucial subroutine of Grover's Algorithm is the Grover iteration, often denoted by $G$:
    \begin{itemize}
      \item Apply Oracle $O:|x\rangle|-\rangle\rightarrow(-1)^{f(x)}|x\rangle|-\rangle$；
      \item Apply Hadamard Gates:$H^{\otimes n}$；
      \item Perform a conditional phase shift($2|0\rangle\langle 0|-I$) on the computer, with every non-zero bases receiving a phase shift of $-1$.
      \item Perform Hadamard transformation $H^{\otimes n}$.
    \end{itemize}
    Note that $H^{\otimes n}(2|0\rangle\langle 0|-I)H^{\otimes n}=2|\psi\rangle\langle\psi|-I$
    One can proof that Grover iteration can be looked as a rotation in the plane spanned vectors which denoted the right answers and the wrong answers.

    Let $\Sigma'_x$ be the sum of all the vectors which indicate a solution to the search problem, $\Sigma''_x$ the rest. Define normalized states:
    $$\begin{array}{ccc}
        |\alpha\rangle & \equiv & \frac{1}{\sqrt{N-M}}\Sigma''_x|x\rangle \\
                       & & \\
        |\beta\rangle & \equiv & \frac{1}{\sqrt{M}}\Sigma'_x|x\rangle
      \end{array}
    $$
    thus the initial state~$|\psi\rangle=\frac{1}{N^{1/2}}\Sigma^{N-1}_{x=0}|x\rangle$~could be represented as
    $$|\psi\rangle=\sqrt{\frac{N-M}{N}}|\alpha\rangle+\sqrt{\frac{M}{N}}|\beta\rangle$$
    The action of Operator~$O$ is~$O(a|\alpha\rangle+b|\beta\rangle)=a|\alpha\rangle-b|\beta\rangle$, which could be looked as perform a reflection in $\alpha\beta-$ plane.
    Similarly Operator~$2|\psi\rangle\langle\psi|-I$ also performs a reflection in~$\alpha\beta-$plane. Thus both two reflections which could be looked as a rotation occur in the $\alpha\beta-$plane. Let $\cos \theta/2=\sqrt{(N-M)/N}$, s.t.$|\psi\rangle=\cos\theta/2|\alpha\rangle+\sin\theta/2|\beta\rangle$, apply the iteration once makes $|\psi\rangle$ become
    $$G|\psi\rangle=\cos\frac{3\theta}{2}|\alpha\rangle+\sin\frac{3\theta}{2}|\beta\rangle$$
    $k$ times use of Grover's Iteration will lead to the following result:
    $$G^{k}|\psi\rangle=\cos\left(\frac{2k+1}{2}\theta\right)|\alpha\rangle+\sin\left(\frac{2k+1}{2}\theta\right)|\beta\rangle$$

    Since $|\psi\rangle=\sqrt{(N-M)/N}|\alpha\rangle+\sqrt{M/N}|\beta\rangle$, we just need to rotate $|\psi\rangle$ $\arccos\sqrt{M/N}$ radians to the one which is parallel to vector $|\beta\rangle$. So repeating $G$ for $R=[\frac{\arccos\sqrt{M/N}}{\theta}]$ times will get $|\psi\rangle$ to within an angle $\theta/2\leq\pi/4$ of $|\beta\rangle$. This is a 'good' state, for people only have to repeat the experiment for expected constant times to get the solution to the problem(Consider geometric probability distribution: $\mathrm{E}[X]=1/(1/2)=2$).

    Apparently~$R\leq\lceil\pi/2\theta\rceil$, suppose~$M\leq N/2$~then we have~$\frac{\theta}{2}\geq\sin\frac{\theta}{2}=\sqrt{\frac{M}{N}}$. Thus, we obtain:
    $$R\leq\left\lceil\frac{\pi}{4}\sqrt{\frac{N}{M}}\right\rceil$$
    in other words we need repeat $G$ for $R=O(\sqrt{N/M})$ times.

    \subsection{Quantum Simulation and Quantum Algorithm}
    Quantum Lattice Celluar Automata(QLCA) and Quantum Gas Automata(QG-A\-) are two familiar ideal models in the research of quantum simulation\cite{mey}~. Meyer, Boghosian\cite{mey,bog1,bog2} have obtained their results respectively by using these models, that is, they
    construct some quantum algorithms which demonstrate exponentially speedup in such models. For Bohosian, the object they tried to simulated is a QGA which obey lattice Boltzman distribution, where arbitrary fields can be concerned with. They have proofed that the complexity of simulation is only related to the dimension of the lattice, but almost has nothing to do with the number of the particles. However, the number of particle always cause exponentially hardness on a classical computer. In fact, Boghosian's results imply that it is almost impossible for a classical computer to simulate one evolution step of a quantum system including dozens of particles.

    We've mentioned that it is the difficulty of quantum simulation that makes people believe quantum mechanics can provide enormous power of computation in the early years.

    Note that in this article we do not care about the hardness of simulations. Generally speaking, the hardness of simulation has nothing to do with the one of computation. For instance, people may find it difficult to simulate some classical celluar automaton according to the given regulations, however once the tedious work has been completed there often exists some more simple methods to produce the series. A typical example is that the regulations of an automata actually cause a circle with a finite period in the series. The same thing can happens to quantum simulations too.

    However, it is important to know that there must exists some cases in which simulations and computations are equivalent. These extreme cases often appears when the length of regulations is near the Kolmogorov complexity(lower bound of description) of a series. Still, strictly speaking, at present no one can proof that polynomially universal unitary operators really cause exponentially difficulty in classical computation. To understand this, just consider an easy but helpful fact that almost all the problems we want to efficiently solved on a quantum computer are in the class BQP, and we have BQP$\subseteq$PSPACE. Unfortunately PSPACE$=$P is not totally impossible. Of course most people don't believe this is true, since this would imply that Shor's Algorithms can be polynomially simulated on a classical computers.

    Now we discuss how to extract a corresponding quantum algorithm from a method of quantum simulation, which is believed to be exponentially faster than any classical one of the same target.

    On a high level, we should do following things:
    \begin{itemize}
      \item Find a family of experiments of quantum mechanics which can be efficiently simulated by quantum computers but are believed to be hardly to simulate and compute by classical computers；
      \item Design a 'good' problem about some non-trivial properties of the last state of the system, which makes quantum computers able to present the answer to the observer quickly.
    \end{itemize}
    Designing the problem is a crucial step. In most cases, though we may have quickly obtained the probabilistic distribution very close to the real experiments, we can not know the whole information in short time. So first we have to ask a question which can be easily verified by any quantum computers containing the whole quantum information of the system.

    For example, we can ask a question such like:
    \begin{itemize}
      \item What the number of the $n_0-$th digit of the probability of a certain system arriving in $\Omega'(\Omega'\subset\Omega)$?
    \end{itemize}

     The problem of this method is that in high dimensional spaces, it is very likely that the probability of the set $\Omega'$ is exponentially close to zero, which actually enables a classical computers to guess zero without running and get the right answer in most cases.

     Now we propose our version: Suppose $\phi$ is the wave function of the system we've simulated and $|\phi(X)|^2,X\subset\Omega$ is the probability of $\vec{x}$ appear in $X$. Try to find two subsets $A,B\subset\Omega$ s.t.$$\frac{3}{7}\leq\frac{|\phi(A)|^2}{|\phi(B)|^2}\leq1$$
     and determine the value of the $n_0-$th digit of $\phi(A)$.

     For the systems which (probabilistic)Turing Machine cannot simulate in polynomial time, the question above is intuitively hard to answer, though up till now no one can proof or disproof it.

     On the other hand, if these systems can be efficiently simulated by quantum computers, repeating following procedure will ensure us to find the answer relatively much faster than any probabilistic Turing Machine of the same aim.
     \begin{defin}\rm \textbf{(Vector of normal vectors $\vec{x}$)}
          $$\vec{x}\equiv\left(\left(\begin{array}{c}
        0 \\
        0 \\
        \vdots \\
        1
      \end{array}\right),
    \left(\begin{array}{c}
        0 \\
        \vdots \\
        1 \\
        0
      \end{array}\right),
      \dots,
      \left( \begin{array}{c}
        1 \\
        \vdots \\
        0 \\
        0
      \end{array}\right)\right)
    $$
     \end{defin}
     \begin{defin}\rm\textbf{(Procedure $P_Q$)}
     $P_Q$(In pseudo-code):\\
     while(find the answer)\\
     \{\\
       . \quad Mid-cut the space $\Omega$ by super-plane whose normal vector is $x_i$.\\
       . \quad Suppose the two spaces is $\Omega_1$ and $\Omega_2$\\
       . \quad if(the condition is satisfied(verifies by testing))\\
       . \quad\{\\
       . \quad\quad halt\\
       . \quad\}\\
       . \quad else\\
       . \quad\{\\
       . \quad\quad$\Omega=\mathrm{min}_{|\phi|}\{\Omega_1,\Omega_2\}$\\
       . \quad\quad$i++$\\
       . \quad\}\\
     \}\\

     \end{defin}

    \section{Conclusions and Future Works}
    We formally proposed the theory of physical computation, define the concepts of resource and complexity. Several examples, including classic mechanics and quantum mechanics, were discussed and analyzed under the framework of physical computation. A technique, which is used to converse a method of quantum simulation into a quantum algorithm, is discussed.

    This is a exciting field, we believe there is more exciting topic to discussed. A very interesting question is: can we find a physical mechanism as the fastest implementation of an arbitrary functions?

    In Sec-III, we talk about the question of calculating the centroid of an object. We thought it is the limitation of dimensions(only three dimensions) hide the advance of the method we mentioned. We conjecture that this method has a excellent counterpart in high dimensional cases. We'll have a try in the (quantum)statistics mechanism.

    In Sec-IV, we talked about quantum simulations and how to construct a clever problem to induced a quantum algorithm. Actually, we conjecture that the problem we construct is a hard one in class $\#P$, for these questions have a counting style. However, we are not sure about whether the designed questions could be in $\#P-hard$ under some specific statistical models. We shall try to work on this in the future.

    We've mentioned that we assume that polynomial qubits is polynomially hard to prepare. However, it is harder to control the qubits as the number of them increase\cite{nc} so far. So one can still conjecture that preparing qubits itself is a ''complicated computing'', and the results up till now can be explained as someone displace the resource consuming procedure, just like DNA Algorithms.

    \section{Acknowledgement}
    The authors are grateful to Haixing Hu, Nan Wu and Jiasen Wu for their useful remarks and suggestions.


    \end{document}